\documentclass[11pt,eqsecnum]{article}
\usepackage{amsmath,amssymb,graphicx} 

\usepackage{psfrag}
\usepackage{subfigure}
\usepackage{tabularx}

\newcommand{\beq}{\begin{equation}}
\newcommand{\eeq}{\end{equation}}
\newcommand{\beqn}{\begin{eqnarray}}
\newcommand{\eeqn}{\end{eqnarray}}
\newcommand{\pa}{\partial}

\newcommand{\appsection}[2]
	{
	\renewcommand{\theequation}{#1-\arabic{equation}}
 	\renewcommand{\thesubsection}{#1.\arabic{subsection}}
	\setcounter{equation}{0}
	\setcounter{subsection}{0}
	\section*{Appendix #1: #2}
	}

\newcommand{\comment}[1]{}
\newcommand{\ZZ}{\mathbb{Z}}

\newcommand{\myfig}[3]{
	\begin{figure}[ht]
	\centering
	\includegraphics[width=#2cm]{#1}\caption{#3}\label{fig:#1}
	\end{figure}
	}

\begin{document}

\def\thistitle{Topological Entanglement Entropy in Chern-Simons Theories
and Quantum Hall Fluids}
\def\thisaddress{Department of Physics, University of Illinois, 1110 W. Green St., 
Urbana IL 61801-3080, U.S.A.}

%

\renewcommand\author[1]{#1}
\rightline{ILL-(TH)-08-03}
\rightline{HIP-2008-04/TH}
\vskip 0.75 cm
\begin{center}
{\Large \textbf{Topological Entanglement Entropy in Chern-Simons Theories
and Quantum Hall Fluids}}
\end{center}
\vskip 0.75 cm
\centerline{{\bf 
\author{Shiying Dong$^1$}, 
\author{Eduardo Fradkin$^1$}, 
\author{Robert G. Leigh$^1$} 
and
\author{Sean Nowling$^{1,2}$}
}}
\centerline{$^1$\it \thisaddress}
\centerline{$^2$\it Department of Mathematics and Statistics and Helsinki Institute of Physics,}\centerline{\it University of Helsinki, P.O. Box 64, 00014 Helsinki, Finland
}


\begin{abstract}
We compute directly the entanglement entropy of spatial regions in Chern-Simons gauge theories in $2+1$ dimensions using surgery. We use these results to determine the universal topological piece of the entanglement entropy for Abelian and non-Abelian quantum Hall fluids. 
\end{abstract}


\section{Introduction}
\label{sec:intro}

The problem of quantum entanglement and its measurement has a long history in quantum mechanics, going back to von Neumann, who introduced the concept of entanglement entropy. The quantum mechanical state of a subsystem $A$ is defined by its reduced density matrix $\rho_A$, obtained by tracing out the information contained in $B$,  the rest of the system. Here $A$ and $B$ are a partition of a larger system which is assumed to be in a pure quantum state. The von Neumann entanglement entropy $S_A$ ($S_B$) of region $A$ ($B$) is defined to be 
\begin{equation}
S_A=-\textrm{tr} \left(\rho_A \ln \rho_A\right)=-\textrm{tr} \left(\rho_B \ln \rho_B \right)=S_B.
\end{equation}
For a quantum mechanical system with a finite (and typically) small number of degrees of  freedom, the von Neumann entropy is a useful and quantitative way to quantify the entanglement encoded in a quantum state.

The entanglement entropy in a quantum field theory, and for that matter in any quantum mechanical system with an infinite (macroscopic) number of degrees of freedom, is in general a complicated non-local quantity whose properties are not well understood. In quantum field theory, interest in the properties of the entanglement entropy  arose in the context of finding a possible explanation of the Bekenstein-Hawking area law of black hole thermodynamics in terms of quantum information concepts.  In a local quantum field theory in $d$ space dimensions, the entanglement entropy for a finite region of linear size $L$ scales with the size of the boundary (``area'') $\left(L/a\right)^{d-1}$ of the region, but with a non-universal ({\it i.e.\/}, dependent on the choice of the ultraviolet cutoff $a$) prefactor without an {\it a priori} relation to any general-relativistic quantities.\cite{srednicki93} Further studies showed that in $1+1$-dimensional conformal quantum field theories this generally non-universal field-theoretic area law reduces to a {\em universal} $\ln \left(L/a\right)$ size dependence with a {\em universal coefficient} equal to $c/3$, where $c$ is the central charge of the conformal field theory\cite{callan-wilczek-1994,calabrese04}. Aside from these important results, little else is known about the behavior of the entanglement entropy in quantum field theory. 

On the other hand, interest in the behavior of the entanglement entropy in condensed matter systems arose in the context of studies of systems near quantum critical points. In that context, the concept of entanglement entropy offers a new perspective to characterize the behavior of quantum critical points from a unique quantum mechanical perspective. However, although the behavior of the entanglement entropy has been studied in a number of interesting quantum critical systems \cite{calabrese04,vidal03,Verstraete06,refael-moore-2004,fradkin-moore-2006,kopp06}, its behavior at generic critical points is not yet well understood. Progress on this problem is of general interest since  deeper understanding of the scaling behavior of the entanglement entropy near quantum critical points will shed light on its general structure in quantum field theory, and vice versa. In addition, this new way of characterizing quantum phase transitions is of interest in the context of current efforts to use such systems  for quantum computing. 

It turns out that the theories for which the concept of entanglement entropy is particularly powerful are {\em topological quantum field theories}. The best understood topological quantum field theory is the Chern-Simons gauge theory in $2+1$ dimensions.\cite{witten89,witten92}  The main purpose of this paper is  to determine the connection between the entanglement entropy of Chern-Simons gauge theory and its topological data. We will work out the entanglement entropy for a general Chern-Simons theory with gauge group $G$ and level $k$ on general spatial topologies. We will then apply our results to the cases of most physical interest.  In addition to their intrinsic interest in topological field theory, our results are  relevant to the study of topological phases of condensed matter systems whose low energy effective field theories are topological quantum field theories. Our results may also have relevance in black hole physics in view of Witten's conjecture on the relation between the  Ba{\~n}ados-Teitelboim-Zanelli  black hole \cite{Banados92} of $2+1$-dimensional gravity and Chern-Simons gauge theory.\cite{witten07} It might also prove interesting to compare these holographic results with a direct computation of the gravitational entanglement entropy using the Chern-Simons description of $2+1$ dimensional gravity.  

Although some of the concepts we discuss here can in principle have wider applicability, in this paper we will only concern ourselves with topological field theories in two space dimensions.  It has been shown~\cite{kitaev-preskill-2006,levin-wen-2006} that for a field theory in two space dimensions that is topological in a limit,  the entanglement entropy for a large simply connected region $A$  of linear size $L$ with a smooth boundary (a subset of an effectively infinite simply connected system) has the form
\begin{equation}
S_A=\alpha L-\gamma ,
\label{eq:Stopo}
\end{equation}
where $\alpha$ is a non-universal coefficient. This form holds provided the linear size $L$ of the region is large compared to any intrinsic length scale of the theory. The universal constant term $\gamma$, known as the {\em topological entropy},
 characterizes the topological state and it is a property of a topological field theory. For a general topological field theory it is given by~\cite{kitaev-preskill-2006,levin-wen-2006}
\begin{equation}
\gamma=\ln {\mathcal D}=\ln \sqrt{\sum_i d_i^2},
\label{eq:effdim}
\end{equation}
where $d_i$ are the quantum dimensions of the quasiparticles (labelled by $i$) of the excitation spectrum associated with  this phase, and $\mathcal{D}$ is  the effective quantum dimension.\cite{Preskill04}

Topological phases in two space dimensions are states of matter which satisfy the following properties. They are ``liquid'' phases, translationally invariant ground states that do not break spontaneously any symmetries of the system. On manifolds with a non-trivial topology ({\it e.g.\/} a torus) the ground states exhibit a non-trivial degeneracy which is robust since it cannot be lifted by the action of any local perturbation. In these phases the excitation spectrum is  gapped.  In the limit of low energies and long distances the wave functions of a set of excitations, vortices of these fluids, exhibit non-local properties that do not depend on the positions of the excitations.
The states of the excitations of a topological state span a topologically protected Hilbert space, and the rate of growth of the dimension of this Hilbert space (as a function of the number of excitations of type $i$) is the quantum dimension $d_i$ of the excitation. In the limit of a vanishing correlation length $\xi \to 0$, the effective field theory of a topological phase   is a topological field theory. The path integral (partition function) of a topological field theory is a topological quantity in the sense that it is independent of the metric of the space. The prototype of a topological field theory is Witten's Chern-Simons gauge theory of the Jones polynomial\cite{witten89,witten92}.

The best known and most studied (both experimentally and theoretically) topological phases in condensed matter are the fractional quantum Hall (FQH) fluids, incompressible phases of two-dimensional electron gases (2DEG) in large magnetic fields
(for a review see Ref.\cite{prange:QHE,dassarma-pinczuk-1997}). Another experimentally accessible candidate for a  topological phase is the superconducting phase of the quasi-two-dimensional strongly correlated oxide Sr$_2$RuO$_4$ 
which appears to be a $p_x+ip_y$ superconductor and is also a topological state.\cite{read-green-2000,stone-chung-2006} Ultra-cold bosonic gases in rotating magnetic traps have also been conjectured to form bosonic analogues of the fractional quantum Hall states.\cite{cooper-wilkin-gunn-2001}

The non-local behavior of the excitations of a topological phase in two space dimensions is closely related to the braiding properties of their world lines which, in turn,  determine the analytic properties of their wave functions. These excitations are generally known as anyons and carry fractional statistics.\cite{leinaas77,wilczek82} Excitations with Abelian fractional statistics are labeled by one-dimensional representations of the braid group and their quantum dimensions $d_i=1$, and their associated Hilbert spaces are one dimensional. Excitations with non-Abelian statistics are labeled by multi-dimensional representations of the braid group, have quantum dimensions $d_i>1$, and their associated Hilbert spaces are multi-dimensional. Such non-Abelian excitations,  and their topologically protected Hilbert spaces, are the basis of the concept of topological quantum computation.\cite{kitaev-2003,freedman-larsen-wang-2002,dassarma07}
    
 In spite of its non-local nature, the topological entanglement entropy of FQH states is actually of direct physical interest, and important for the characterization of quantum Hall interferometers. Quantum interferometers of FQH fluids are devices which in principle can detect the fractional statistics of the FQH quasiparticles.\cite{chamon-1997,fradkin-nayak-tsvelik-wilczek-1998,slingerland-bais-2001,kim-lawler-vishveshwara-fradkin-2005,dassarma-2005,stern-halperin-2006,bonderson-kitaev-shtengel-2006,ardonne-kim-2007,feldman-2007}  It has recently been shown~\cite{fendley-fisher-nayak-2007a,fendley-fisher-nayak-2007b,fendley-fisher-nayak-2007c} that a quantum point contact in a non-Abelian quantum Hall state actually can act as a quantum disentangler. This is possible because point contacts in a quantum Hall system essentially consist of places where the tunneling matrix elements between the edge states is non-vanishing. From the point of view of the bulk FQH state this is a non-local connection which disrupts its quantum correlations. Remarkably, Fendley and coworkers\cite{fendley-fisher-nayak-2007a,fendley-fisher-nayak-2007b,fendley-fisher-nayak-2007c} found that the change of the entanglement entropy induced by the point contact is equal to the change in the Affleck-Ludwig entropy~\cite{affleck-ludwig-1991} of the edge states of the FQH fluid. Thus, the topological entropy is a quantity of interest in the present effort to develop interferometers for quantum Hall quasiparticles.

  In this paper we investigate the universal properties of the entanglement entropy of Chern-Simons gauge theory for a general compact gauge group at an arbitrary level $k$. We give explicit results for $SU(2)_k$ and for several coset conformal filed theories of interest for applications.  We use our results to compute the entanglement entropy of fractional quantum Hall topological fluids by computing this quantity directly at the level of the effective topological field theory. By working directly in the topological limit we obtain directly the $O(1)$ term in the entropy, the topological entropy. All other size-dependent terms, including the ``area term'', become zero in this limit. Naturally, size-dependent terms will arise if (irrelevant) corrections to the topological action, such as Maxwell/Yang-Mills type terms, were to be included.
 Throughout this paper we will use the path integral representation of Chern-Simons theory. To this end we will adapt the seminal results of Witten\cite{witten89,witten92} for the partition functions of Chern-Simons gauge theories to the computation of the topological entanglement entropy. We use the standard ``replica'' approach to compute the entropy\cite{Holzhey94,calabrese04}. This requires to understand what is the 3-manifold resulting from gluing $n$ copies of the system in a  suitable fashion \cite{calabrese04}, needed to compute the entropy for a number of cases of interest. The key aspect of our approach is the identification of a suitable configuration of Wilson loops for each case of interest and to compute it by reducing it to already known cases by using surgeries. Alternatively, it is also possible to use a more conventional approach using the wave function of the Chern-Simons gauge theory\cite{witten92}. This approach is technically more involved and will only be discussed briefly.  
 
We consider first the case of  a surface of genus zero (a sphere or a disk).\footnote{As this manuscript was being completed we became aware of the very recent work of K. Hikami \cite{hikami-2007}, who calculated entanglement entropies in $SU(2)_k$ theories on the sphere using a skein relation approach. Our results agree with Hikami's where they overlap.}  For a  simply connected region  we compute the topological entropy of the vacuum (ground) state of the Chern-Simons theory on a sphere, and recover the result obtained by Kitaev and Preskill\cite{kitaev-preskill-2006}, and of Levin and Wen\cite{levin-wen-2006}, {\it i.e.,\/} eqs.\eqref{eq:Stopo} and \eqref{eq:effdim}. Next we generalize these results to the case of manifolds with non-vanishing genus (mainly a torus), which have a finite-dimensional topologically protected degenerate vacuum sector. Here we also consider the entanglement of multiply connected regions. We find that the entanglement entropy of a simply connected region is independent of the genus of the manifold, even if the vacuum sector is degenerate. In the case of a multiply-connected region, we find that the entropy scales linearly with the number of components of the observed region only if the vacuum sector is non-degenerate (genus zero). However, if the manifold has a non-vanishing genus, and thus has a degenerate vacuum sector,  in general  the entanglement entropy of multiply-connected regions is different for different states in the vacuum Hilbert space. In other words, the entanglement entropy, aside from the purely topological entropy,  has additional contributions that depend on the choice of state {\it i.e.\/}, on the coefficients of its wave function and on the representation carried by the state. We also compute the entanglement entropy of a simply connected region with several quasiparticles, {\it i.e.\/} operators represented by punctures carrying non-trivial representations. In this case we find that the entanglement entropy generally depends on the conformal blocks in which these operators can fuse, and hence depend explicitly on the structure of the fusion rules. These results indicate that measurements of the entanglement entropy can, in principle, be used to determine the full structure of the underlying effective topological theory. Finally, we apply these results to the computation of the entanglement entropies of fractional quantum Hall fluids. Here we derive the modular $\mathcal{S}$-matrices for several coset CFTs needed to compute the entanglement entropies for non-Abelian FQH states.
  
This paper is organized as follows.  In Section \ref{sec:CS} we set up the calculation of the topological entanglement entropy $\gamma$ in Chern-Simons gauge theories. In Subsection  \ref{subsec:preliminaries} we show how the computation of the entropy can be carried out using the methods developed by Witten \cite{witten89} and use them to compute the entropy for the simplest cases, a simply connected region on a sphere (Subsection \ref{subsec:S2-1comp}) and a torus (Subsection \ref{subsec:T2-1comp}), and multiply-connected regions on a sphere (Subsection \ref{subsec:S2-2comp}) and on a torus (Subsection \ref{subsec:T2-2comp}). In Section \ref{sec:punctures} we present a calculation of the entanglement entropy in the presence of punctures ({\it i.e.\/} quasiparticles) carrying different representation labels. Here we discuss the case of four  quasiparticles on $S^2$ (Subsection \ref{subsec:S2-4}) and  discuss two different cases, paired and not paired, as well as three quasiparticles on $S^2$ (Subsection \ref{subsec:S2-3}).
In Section \ref{sec:fqhe} we present the calculation of the entanglement entropy for both Abelian $U(1)_k$ (Subsection \ref{subsec:U1}) and non-Abelian FQH fluids, in terms of coset Chern-Simons gauge theories (Subsections \ref{subsec:coset1} and \ref{subsec:coset2}). Section  \ref{sec:conclusions} is devoted to the conclusions. The hydrodynamic, Chern-Simons, description of the FQH fluids (both Abelian and non-Abelian)  is summarized in Appendix A, and the calculation of the modular $\mathcal{S}$-matrix in Appendix B.

\section{Entanglement entropy and Chern-Simons gauge theory}
\label{sec:CS}

\begin{figure}[ht]
  \hfill
  \begin{minipage}[t]{.45\textwidth}
	 \includegraphics[width=6cm]{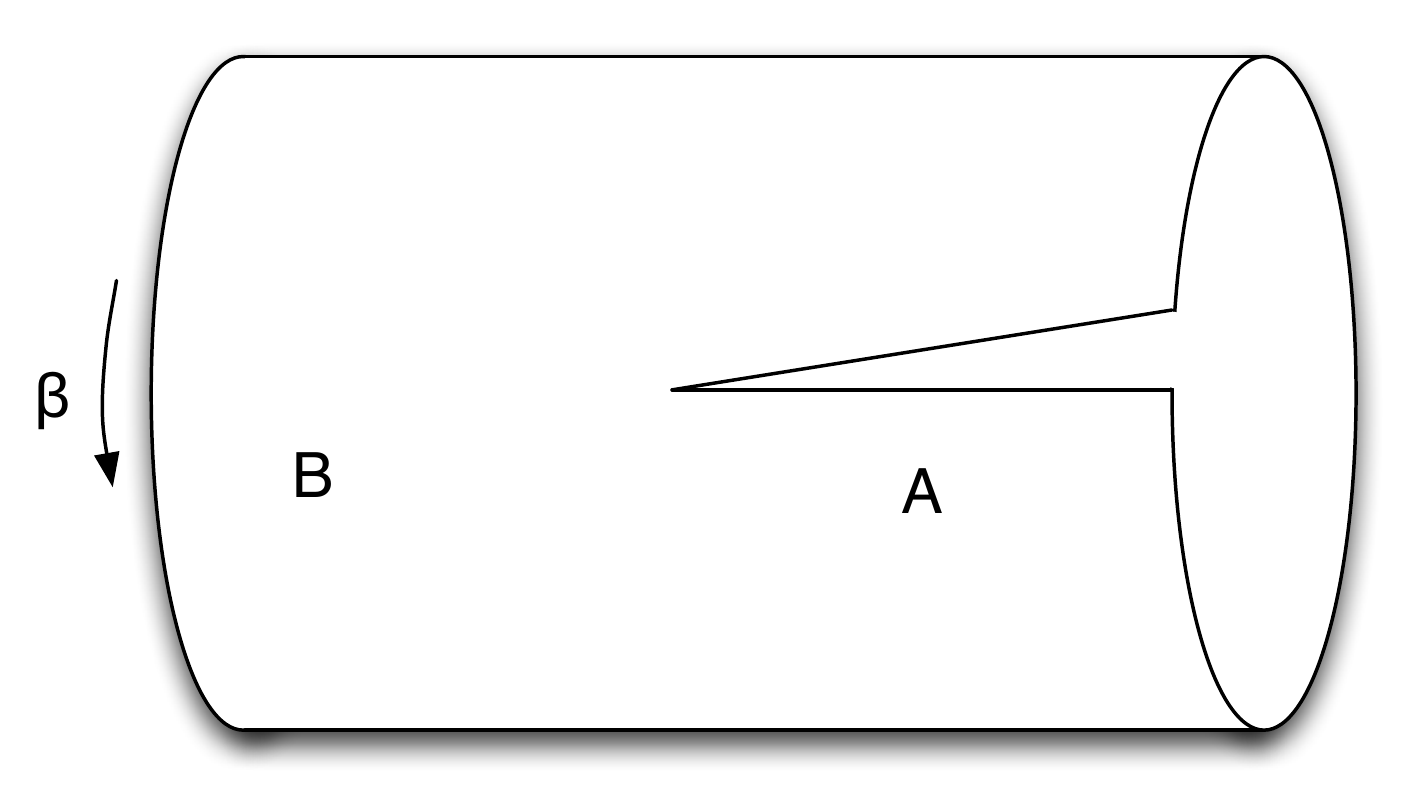}
		\caption{Conceptual picture of $\rho_A$. The trace over B corresponds to gluing $\tau=0$ to $\tau=\beta$ in the B region, leaving a cut open in the A region.}
		\label{fig:Cut_Tube_No_Fields}
  \end{minipage}
  \hfill
  \begin{minipage}[t]{.45\textwidth}
	 \includegraphics[width=6cm]{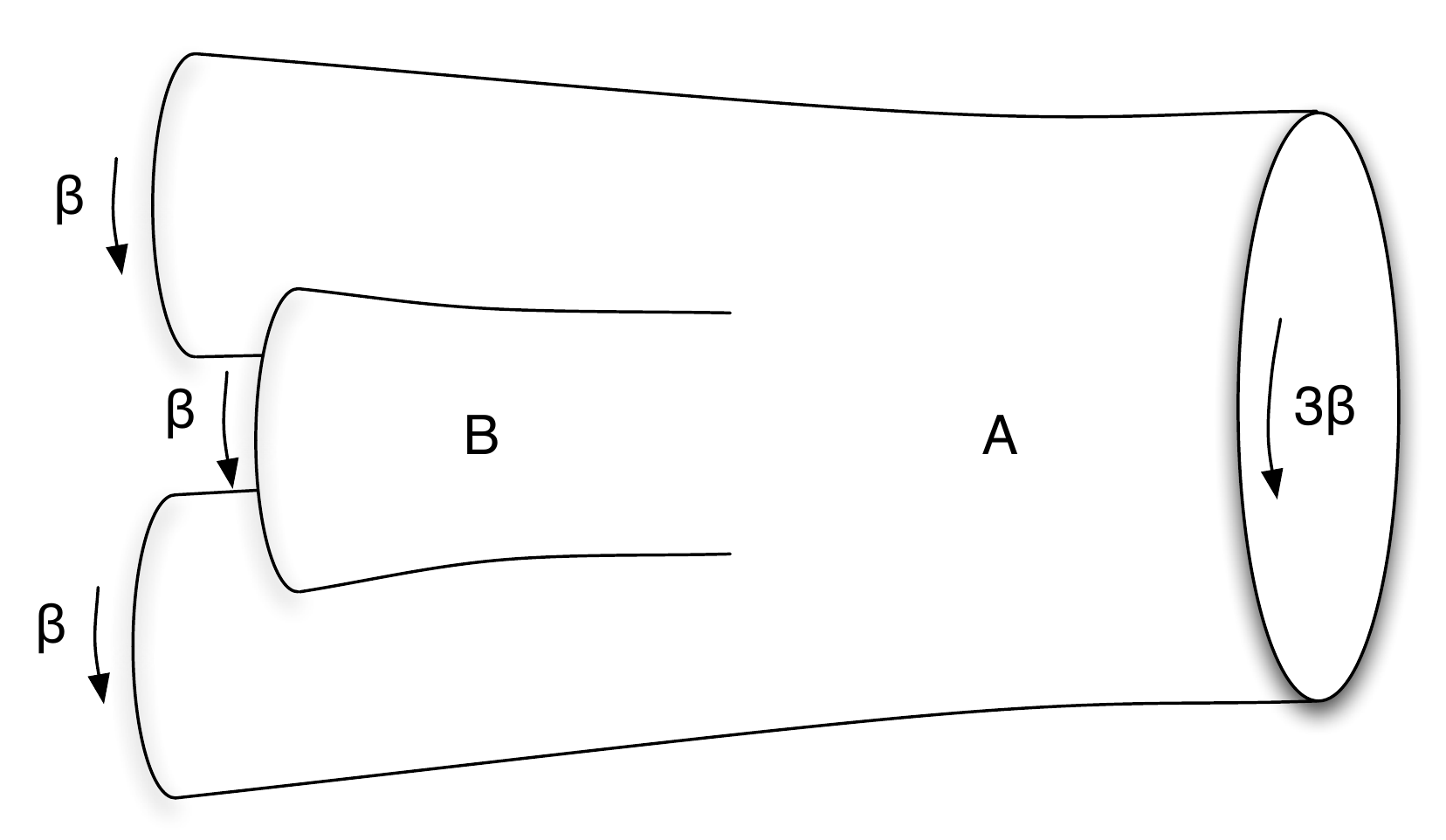}
		\caption{$\textrm{tr}\rho_A^3$ is obtained by gluing three copies of the diagram in
		Fig. \ref{fig:Cut_Tube_No_Fields} back to back along the cut in the A region.}
		\label{fig:3_Glued_Tubes}
  \end{minipage}
  \hfill
\end{figure}

In this paper, we will consider entanglement entropy in Chern-Simons theory in three dimensions. As we will argue in what follows, the entanglement entropy may be obtained by computing a Chern-Simons path integral on certain 3-geometries, which we systematically obtain through a `gluing' procedure. To see what this procedure should be, we review here the conceptually simpler case of a scalar field theory. 
The 2-dimensional case was described by Calabrese and Cardy.\cite{calabrese04}
Consider a spatial domain which we slice into two, labeled $A$ and $B$. These regions may be connected or not, simply connected or not. We label the interface between $A$ and $B$ by $I=\partial A=\partial B$, which may in general consist of several components. We label the degrees of freedom as $\phi$. To make the discussion more straightforward, we will consider the corresponding finite temperature density matrix
\beqn
\rho\left[ \{\varphi_0(\vec x)\}, \{\varphi_\beta (\vec x)\}\right] &=& \frac{1}{Z(\beta)}
\langle  \{\varphi_0(\vec x)\} | e^{-\beta \hat H} | \{\varphi_\beta(\vec x)\}\rangle\\
&=& \int \prod_{\vec x,\tau}\left[ d\phi(\vec x,\tau)\right] e^{-S_E} \prod_{\vec x} \delta\left[ \phi(\vec x,0)-\varphi_0(\vec x)\right] \delta\left[ \phi(\vec x,\beta)-\varphi_\beta(\vec x)\right],
\eeqn
where we specify the state by a spatial configuration at $\tau=0,\beta$.
In this language, a trace is obtained by path integration over $\varphi_0$ and $\varphi_\beta$. Having split the spatial domain into pieces, we may then obtain the reduced density matrix $\rho_A$ by tracing over $B$,
\beq
\rho_A\left[ \{\varphi_0(\vec x)\}, \{\varphi_\beta (\vec x)\}\big| \vec x\in A\right] 
=\int \left(\prod_{\vec x\in B} \left[ d\varphi_0(\vec x)d\varphi_\beta(\vec x)\right]\delta\left[\varphi_0(\vec x)-\varphi_\beta (\vec x)\right]\right)
\rho\left[ \{\varphi_0(\vec x)\}, \{\varphi_\beta (\vec x)\}\right].
\eeq
The entanglement entropy will be obtained by a replica trick,
\beq
\label{deftrace} S_A =-\textrm{tr}\rho_A\ln \rho_A = -\frac{d}{dn} \textrm{tr}\rho_A^n \Big|_{n=1}
\eeq
(we expect that $\textrm{tr}\rho_A^n$ will have a unique analytic continuation in $n$ for $n\geq 1$).
Finally, $\textrm{tr}\rho_A^n$ is obtained by taking $n$ copies of $\rho_A$ and `gluing' them together appropriately
\beqn
\textrm{tr} \rho_A^n=\int \prod_{k=1}^n\left\{
\prod_x [d\varphi_0^{(k)}(\vec x)d\varphi_\beta^{(k)}(\vec x)]
\prod_{x\in A} \delta \left[\varphi_0^{(k)}(\vec x)-\varphi_\beta^{(k+1)}(\vec x)\right]
\prod_{x\in B} \delta \left[\varphi_0^{(k)}(\vec x)-\varphi_\beta^{(k)}(\vec x)\right]
\right.\nonumber\\ \left.
\rho\left[\{\varphi_0^{(k)}(\vec x)\} , \{\varphi_\beta^{(k+1)}(\vec x)\}\right]
\right\}.
\eeqn
This path integral may be interpreted as a scalar field theory defined on a glued manifold, of the form displayed in Figures \ref{fig:Cut_Tube_No_Fields},\ref{fig:3_Glued_Tubes}.

Now, in fact we are not really interested in the entanglement entropy obtained from the finite temperature density matrix. Instead, we would like to pick a pure state (of the whole system); this may be achieved here by taking $\beta\to\infty$. In this limit, the system will project down to the ground state. There is a subtlety here, that will in fact arise in the Chern-Simons theory (or generically in any topological field theory), in that the ground state need not be unique. Thus, the procedure we have outlined is not powerful enough to select a particular degenerate pure state. In the particular case of Chern-Simons theory, we will take the above construction as indicative that we should consider the Chern-Simons path integral on the glued geometry; in this construction, it is clear how to make the choice of a pure state, as we will detail a little later.

In the case of Chern-Simons theory, we can formally perform the above construction by identifying the Chern-Simons wave functional; assuming holomorphic factorization, this may be written as a WZW path integral\footnote{Written in this form, the gauge field measure includes a factor $e^{\frac{k}{2\pi}\int \textrm{tr} \bar A A}$.}
\beq\label{cswave}
\langle \bar B_i|\langle \bar A_i|\Psi\rangle \sim \int [dg_{A,i} dg_{B,i}] e^{-kI_A(g_{A,i})-kI_B(g_{B,i})-\frac{k}{2\pi}\int_{\Sigma_A} \textrm{tr}\bar A_i g_{A,i}^{-1}\pa g_{A,i} - 
\frac{k}{2\pi}\int_{\Sigma_B} \textrm{tr}\bar B_i g_{B,i}^{-1}\pa g_{B,i}}.
\eeq
Here, $I(g)$ is a WZW action.
The expression \eqref{cswave} should be interpreted as a sum over histories with a spatial section of fixed topology. Here, we have split the integral into contributions of fields in regions A and B.
Formally, $\textrm{tr}\rho_A^n$ may then be constructed  by gluing together suitable such factors,
\beq
\int\prod_{k=1}^n [d\mu(A_k)d\mu(B_k)] 
\langle \bar B_1|\langle \bar A_1|\Psi\rangle
\langle \Psi| B_1\rangle |A_2\rangle
\langle \bar B_2|\langle \bar A_2|\Psi\rangle
\langle \Psi| B_2\rangle |A_3\rangle
\ldots
\langle \bar B_n|\langle \bar A_n|\Psi\rangle
\langle \Psi| B_n\rangle |A_1\rangle.
\eeq
This can be interpreted as Chern-Simons theory on a glued 3-geometry. This 3-geometry will be determined by a choice of spatial topology (that is, a Riemann surface $\Sigma$ of genus $g$) and a choice of cutting into $A$ and $B$ regions. The original 3-geometry (before gluing) may be mapped to a solid geometry $\tilde \Sigma$, consisting of $\Sigma$ and its interior.
 It is well known that the Hilbert space ${\cal H}_\Sigma$ of this theory is accounted for by the appropriate conformal blocks of the corresponding WZW conformal field theory. For example, for the sphere $S^2$ there is a unique state, while for the torus $T^2$, the various degenerate states may be obtained by placing Wilson lines in representation $R$ along the centre of the solid torus. For higher genus, we can consider the various conformal blocks directly. Thus, the choice of pure state is made here by a choice of conformal block (equivalently, for the torus, a choice of Wilson loop).

\subsection{Modular Properties}
\label{subsec:preliminaries}

Indeed, it is well known that the states of a Chern-Simons theory are accounted for by the conformal blocks of a conformal field theory. As a result, the Chern-Simons states may be identified with characters. The modular ${\cal S}$-matrix of the conformal field theory will then enter in calculations of Wilson loop observables in the Chern-Simons theory, as was exemplified by Witten\cite{witten89}. Consequently, the entanglement entropy will generically depend on matrix elements of the modular ${\cal S}$-matrix. Given a set of characters $\chi(\tau)$ of a CFT, one writes
\beq
\chi(-1/\tau)={\cal S}\chi(\tau),
\eeq
which should be understood as matrix multiplication. The characters are indexed by a set of quantum numbers, which in the case of affine algebras can be taken to be representations. In this paper, we will not need to specify the precise Chern-Simons theory (that is the calculations are valid in general), although physical applications will imply a choice. An example of interest is $\widehat{SU(2)}_k$ WZW, in which representations $\hat R_j$ are labelled by a half-integer $j=0,1/2,...,k/2$, and
\beq
\chi^{(k)}_j(-1/\tau)=\sum_{j'}{{{\cal S}^{(k)}}_j}^{j'}\chi^{(k)}_{j'}(\tau),
\eeq
where
\beq
{{{\cal S}^{(k)}}_j}^{j'} = \sqrt{\frac{2}{k+2}}\sin \left[\frac{\pi (2j+1)(2j'+1)}{k+2}\right].
\label{eq:Smatrix-SU2}
\eeq
More generally, the ${\cal S}$-matrix is assumed to be unitary. Thus, we have
\beq
{{\cal S}_i}^{j} {({\cal S}^\dagger)_{j}}^{k}={\delta_i}^{k},
\eeq
while applying ${\cal S}$ twice corresponds to charge conjugation\footnote{The notation $\bar j$ refers to the conjugate representation to that labelled by $j$.}
\beq
{({\cal S}^2)_i}^j = {C_i}^j={\delta_i}^{\bar j}.
\eeq
Also of importance are the fusion rules for two representations ${\hat R}_i \times {\hat R}_j$. The multiplicity of representation ${\hat R}_k$ in the fusion is denoted by ${N_{ij}}^k$ which are related to the modular $\mathcal{S}$-matrix by the Verlinde formula
\beq
{N_{ij}}^k=\sum_\ell \frac{{{\cal S}_i}^{\ell}{{\cal S}_j}^{\ell}{({\cal S}^{-1})_\ell}^{k} }
{ { {\cal S}_0}^{\ell}}
\label{eq:verlinde}
\eeq

The {\it quantum dimension} is defined as
\beq
d_j=\frac{{{\cal S}_0}^j}{{{\cal S}_0}^0}.
\eeq
For $\widehat{SU(2)}_k$,  the quantum dimensions are
\beq
d_j=\frac{ \sin\left(\frac{\pi(2j+1)}{k+2}\right)}{\sin\left(\frac{\pi}{k+2}\right)}.
\eeq
In Section \ref{sec:fqhe} we give generalizations of these formulae to other CFTs of interest.

We note that the unitarity condition implies
\beq
({{\cal S}_0}^0)^{-1} =\sqrt{\sum_j |d_j|^2}={\cal D}.
\eeq

In the Chern-Simons theory, we will be led to evaluate the partition function on various 3-geometries with Wilson loops. These can be systematically computed by a series of ``surgery'' operations\cite{witten89}. The result of these computations is that the partition functions depend on various matrix elements of modular matrices. For example,
\beq\label{s3z}
Z(S^3,\hat{R}_j)={{\cal S}_0}^j,
\eeq
where the notation on the left means the Chern-Simons partition function on $S^3$ with a Wilson loop in representation $\hat R_j$ (where $j$ is an index labeling representations). 

Another basic result that we will use repeatedly applies to a 3-manifold $M$ which is the connected sum of two 3-manifolds $M_1$ and $M_2$ joined along an $S^2$. We have (Eq.(4.1) in \cite{witten89})
\beq\label{surgconnsum}
Z(M)\cdot Z(S^3)=Z(M_1)\cdot Z(M_2).
\eeq
This result relies crucially on the fact that the Hilbert space for $S^2$ is one dimensional. Similarly, using the same reasoning, we can deduce that if $M$ is $M_1$ and $M_2$ joined along $n$ $S^2$'s,
\beq\label{connsumZ}
Z(M) =\frac{Z(M_1)\cdot Z(M_2)}{Z(S^3)^n}.
\eeq
To demonstrate this, we note that the path integral on a connected sum of $M_1$ and $M_2$ (that is, $M_1$ and $M_2$ joined through $S^2$'s) can be thought of as the overlap of states $|\chi_1\rangle$ and $|\chi_2\rangle$ each defined on the interface $S^2$, $Z(M_1+M_2)=\langle\chi_1|\chi_2\rangle$.
Because the Hilbert spaces involved are {\it one-dimensional}, we can insert a state in between which topologically corresponds to capping off the connection (that is, we sew in half of a 3-ball onto $M_1$ and half of a 3-ball (with opposite orientation) onto $M_2$). Thus,
\beq
Z(M_1+M_2)=\frac{\langle\chi_1|\chi_3\rangle\langle\chi_3|\chi_2\rangle}{\langle\chi_3|\chi_3\rangle}
=\frac{Z(M_1)\cdot Z(M_2)}{Z(S^3)},
\eeq
the latter equality obtaining because $\langle\chi_3|\chi_3\rangle$ corresponds to a 3-sphere.
We can repeat this construction, capping off each join, to obtain the more general result Eq.\eqref{connsumZ}. In the following, this is the general surgery operation that we employ repeatedly.

\subsection{$S^{2}$ with one A-B interface}
\label{subsec:S2-1comp}

Let us begin with the simplest case, in which the spatial topology is a 2-sphere. The Hilbert space on $S^2$ is one dimensional, and so there is only one choice of state. The 3-geometry is the 3-ball shown in Fig. \ref{fig:S2single}.
 \myfig{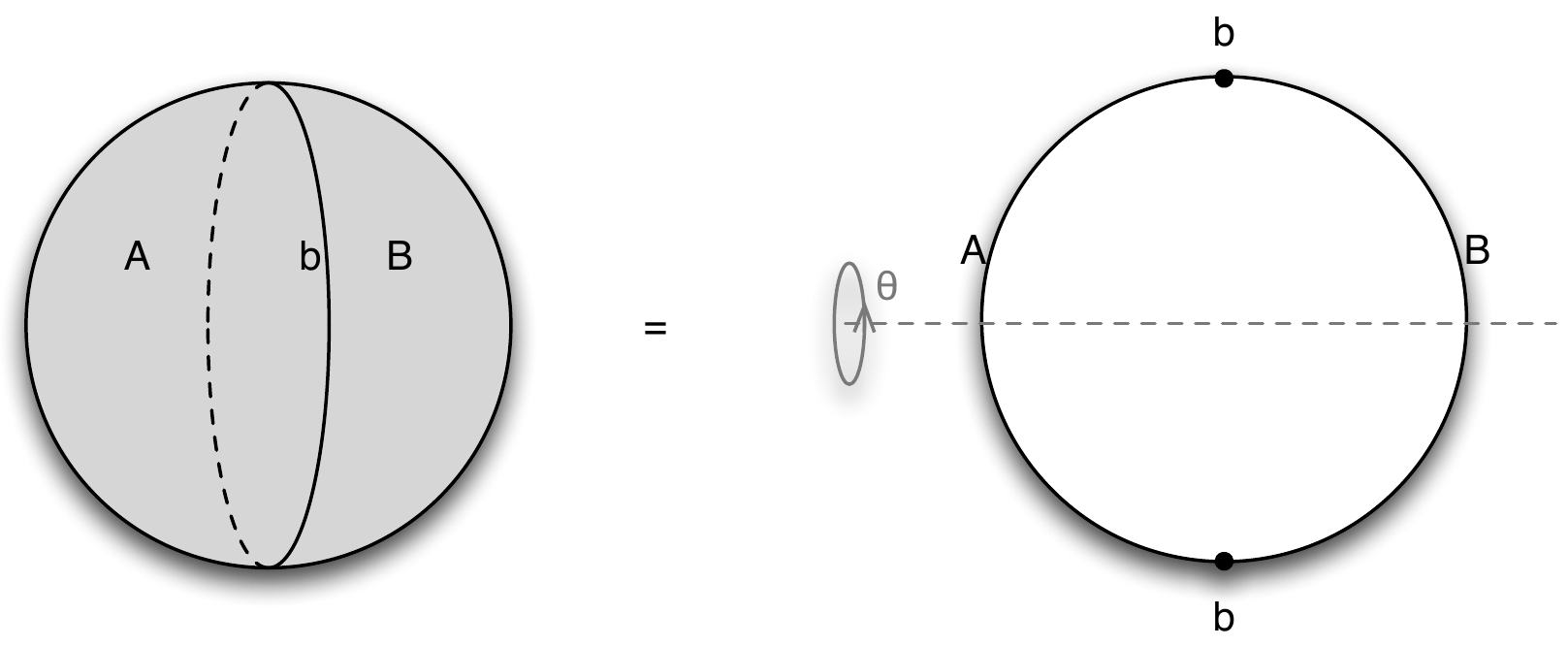}{9}
{Shading implies a solid 3-ball. With a one-component interface, the A and B regions are disks. It is useful in the following constructions to view the 3-ball as a disk rotated about an axis passing through the origin, as shown at right.}
If we take the A and B regions to be connected, then they are disks. To construct $\textrm{tr}\rho_A^n$,  we glue $2n$ such pieces together. In Fig. \ref{fig:S2double}, we show how to systematically perform this gluing. We have drawn the $n=2$ case explicitly, but it is not hard to generalize to higher $n$.  
 \myfig{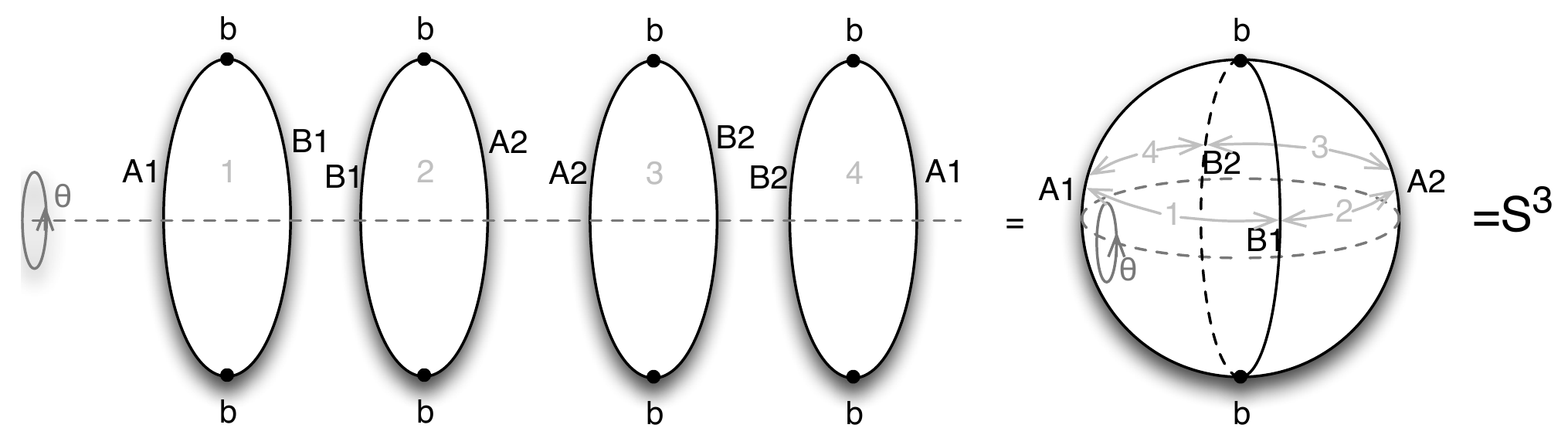}{12}
{For spatial topology $S^2$ with one interface component, we explicitly show the construction of $\textrm{tr}\rho_A^2$. The overall manifold is generated by four pieces of disks glued together one after another and rotated along the same axis as in Fig. \ref{fig:S2single}.}
In the figure, we have used 1 and 2 to label $|\psi\rangle_{1}$ and $\langle\psi |_{1}$, 3 and 4 for $|\psi\rangle_{2}$ and $\langle\psi |_{2}$. The four slices form four 3-balls (or as shown, rotated disks); when glued together to form $\textrm{tr}\rho_A^n$, we find an $S^{2}$, rotated about the axis, which has the topology $S^{3}$. One can easily check that for higher $n$, we obtain the same result, the $S^2$ being obtained by sequentially gluing $2n$ disks. Thus we have the normalized trace,
\beq
\frac{\textrm{tr}\rho^{n}_{A(S^{2},1)}}{\left(\textrm{tr}\rho_{A(S^{2},1)}\right)^{n}}=\frac{Z(S^{3})}{(Z(S^{3}))^{n}}=(Z(S^{3}))^{1-n}=({{{\cal S}_0}^0})^{1-n},
\eeq
where we have used formula \eqref{s3z} given above. Finally, using eq. \eqref{deftrace}, we obtain
\beq
S_A^{(S^2,1)} = \ln{{{\cal S}_0}^0}.
\eeq
This result applies to any Chern-Simons theory. Since ${\mathcal{S}_0}^0=1/\mathcal{D}$, we recover the known result~\cite{kitaev-preskill-2006,levin-wen-2006} for the topological entropy for a simply connected region of a sphere $S^2$ (or a disk) in terms of the effective quantum dimension $\mathcal{D}$:
\beq
S_A^{(S^2,1)}=\ln {\mathcal{S}_0}^0=-\ln \mathcal{D}.
\eeq
In the case of $U(1)_m$, we have
\beq
{{{\cal S}_0}^0}= \frac{1}{\sqrt{\sum_j |d_j|^2}} = \frac{1}{\sqrt{m}},
\eeq
and hence
\beq
S_A^{(S^2,1)} = -\ln \sqrt{m}.
\eeq
For $\widehat{SU(2)}_k$,  we obtain
\beq
S_A^{(S^2,1)} = \ln \left(\sqrt{\frac{2}{k+2}}\sin \frac{\pi}{k+2}\right).
\eeq
For $k=1,2,3$, this evaluates to 
\beq
S_A = -\ln\sqrt{2},\ \ \  -\ln 2,\ \ \  -\ln\left(\sqrt{\sqrt{5}+5}\right),
\eeq
respectively.  For applications of these formulae to physical models, see Section \ref{sec:fqhe}.

\subsection{$T^{2}$ with one component A-B interface}
\label{subsec:T2-1comp}

The Hilbert space on $T^{2}$ is isomorphic to the space of integrable representations $\hat R_j$ of the Kac-Moody algebra. These states are generated by doing the path integral on a solid torus with Wilson loop in representation $\hat R_j$ lying along the non-contractible loop at the centre. We will consider first a slicing of the torus into A and B regions such that there is a single connected interface component. 
%
%
To define the entanglement entropy, we must choose a pure state, and here we have a choice. To begin, let us first choose the trivial representation (equivalent to no Wilson loop). 
It is useful to consider the solid torus as a solid ball with a handle attached in the A region,
since we have already studied the solid ball in the previous subsection, and to compute $\textrm{tr}\rho_A^n$ here, we need to follow that analysis and also keep track of the gluing of the extra toroidal fixtures.
Note that a solid torus can be thought of as $D_2\times S^1$, and two copies glued together (with opposite orientations) gives an $S^{2}\times S^{1}$. The result of the gluing for $n=2$ is shown in Fig. \ref{fig:T2double}. 
 \myfig{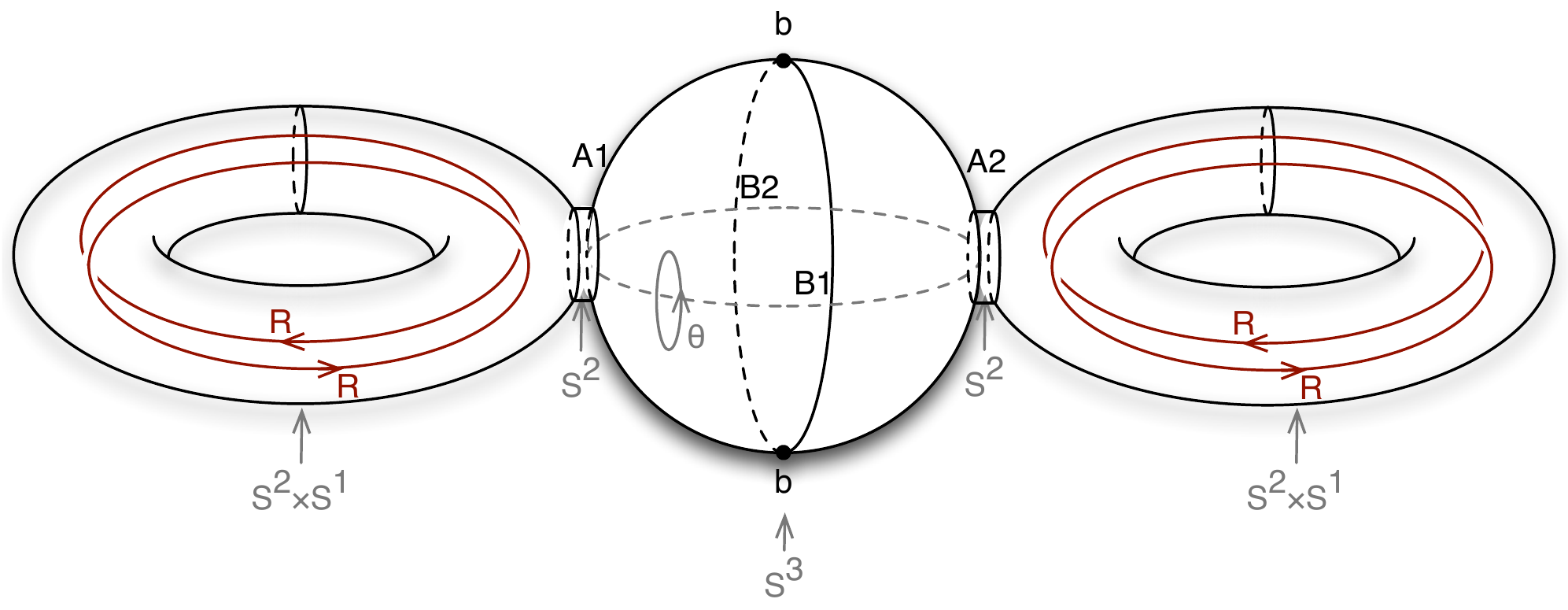}{10}{The space $X_{n=2}$, obtained by gluing $n=2$ copies of the space shown in Fig. \ref{fig:T2single}. The result is an $S^3$ joined to two copies of $S^2\times S^1$ along $S^2$'s. For general $n$, the glued geometry $X_n$ consists of an $S^3$ joined in this way to $n$ $S^2\times S^1$'s.}
Thus the resulting manifold will be the connected sum of an $S^3$ and $n$ $S^{2}\times S^{1}$'s joined along $n$ $S^{2}$'s.  
Thus, applying Eq.\eqref{connsumZ} for $M_1=S^3$ and $M_2$ $n$ disjoint copies of $S^2\times S^1$, we obtain
\beq
\frac{\textrm{tr}\rho^{n}_{A(T^{2},1)}}{\left(\textrm{tr}\rho_{A(T^{2},1)}\right)^{n}}=
\frac{1}{Z(S^2\times S^1)^n}\frac{Z(S^3)Z(S^2\times S^1)^{n}}{Z(S^3)^{n}}=
(Z(S^{3}))^{1-n}=({{\cal S}_0}^0)^{1-n}.
\eeq
The first factor after the first equal sign comes from the normalizing factor (for $n=1$, the topology is just $S^2\times S^1$).
We note that this result coincides with the $S^2$ result. As we shall see, the commonality of these two examples is that the interface is the same; the topology of the A and B regions themselves does not contribute.

It is simple to repeat this construction for other pure states, that is including a Wilson loop in representation $\hat R_j$ inside the solid torus. This Wilson loop is as shown in Fig. \ref{fig:T2single}. In the gluing above, we will now have a $D_2\times S^1$ with Wilson loop in representation $\hat R_j$ glued to a $D_2\times S^1$ of opposite orientation with Wilson loop in representation $\hat{\overline{R}}_j$ (the conjugate state), as indicated in Fig. \ref{fig:T2double}. Thus, we have
\beq
\frac{\textrm{tr}\rho^{n}_{A(T^{2},\hat{R}_j)}}{\left(\textrm{tr}\rho_{A(T^{2},\hat{R}_j)}\right)^n}=\frac{1}{Z(S^{2}\times S^{1},\hat{R}_j,\hat{\overline{R}}_j)^{n}}
\frac{Z(S^3)Z(S^{2}\times S^{1},\hat{R}_j,\hat{\overline{R}}_j))^{n}}
	{Z(S^3)^n}
=Z(S^{3})^{1-n}=({{\cal S}_0}^0)^{1-n}.
\eeq
In fact this result can be generalized further, to {\it any} pure state $|\psi\rangle=\sum_{j}\psi_{j}|\hat{R}_j\rangle$,
\beqn
\frac{\textrm{tr}\rho^{n}_{A(T^{2},\psi)}}{\left(\textrm{tr}\rho_{A(T^{2},\psi)}\right)^n}&=&
\frac
	{\sum_{j_1,j_2,...}\psi_{j_1}\psi_{j_2}^{*}\psi_{j_2}\psi_{j_3}^{*}\dots\psi_{j_n}\psi_{j_1}^{*}Z(X_n,\hat{R}_{j_1},\hat{\overline{R}}_{j_1},...)}
	{\left(\sum_{j}|\psi_{j}|^{2}Z(S^{2}\times S^{1},\hat{R}_j,\hat{\overline{R}}_j)\right)^{n}},
\eeqn
where we have denoted the glued 3-geometry as $X_n$;
the Wilson loops $\hat R_j$ and $\hat{\overline{R}}_j$ are located along the $j^{th}$ toroidal fixture. Performing surgeries as in Eq.\eqref{connsumZ} gives
\beq
\frac
	{\textrm{tr}\rho^{n}_{A(T^{2},\psi)}}
	{\left(\textrm{tr}\rho_{A(T^{2},\psi)}\right)^n}=
\frac
	{\sum_{j_1,j_2,...}\psi_{j_1}\psi_{j_2}^{*}\psi_{j_2}\psi_{j_3}^{*}\dots\psi_{j_n}\psi_{j_1}^{*}\prod_k Z(S^{2}\times S^{1},\hat{R}_k,\hat{\overline{R}}_k)\ Z(S^{3})^{1-n}}
	{\left(\sum_{j} |\psi_{j}|^{2} Z(S^{2}\times S^{1},\hat{R}_j,\hat{\overline{R}}_j)\right)^{n}}=
({{\cal S}_0}^0)^{1-n}.
\eeq
 So we conclude that, at least for this slicing into A and B regions, the entanglement entropy is insensitive to which degenerate pure state we consider. This statement is not generally true, as we will see.

\subsection{$S^{2}$ with two-component AB interface}
\label{subsec:S2-2comp}

Next let's study the case where A and B meet at an interface with two components.
%
%
For $S^{2}$, the only distinct choice is to have two disconnected B regions. 
If we think of this as two solid 3-balls joined together, then we can do the gluing for each 3-ball separately, and then account for the joining. The result is a pair of $S^3$'s, joined  along $n=2$ $S^2$'s, as indicated in Fig. \ref{fig:S2-2double} for the $n=2$ case.
 \myfig{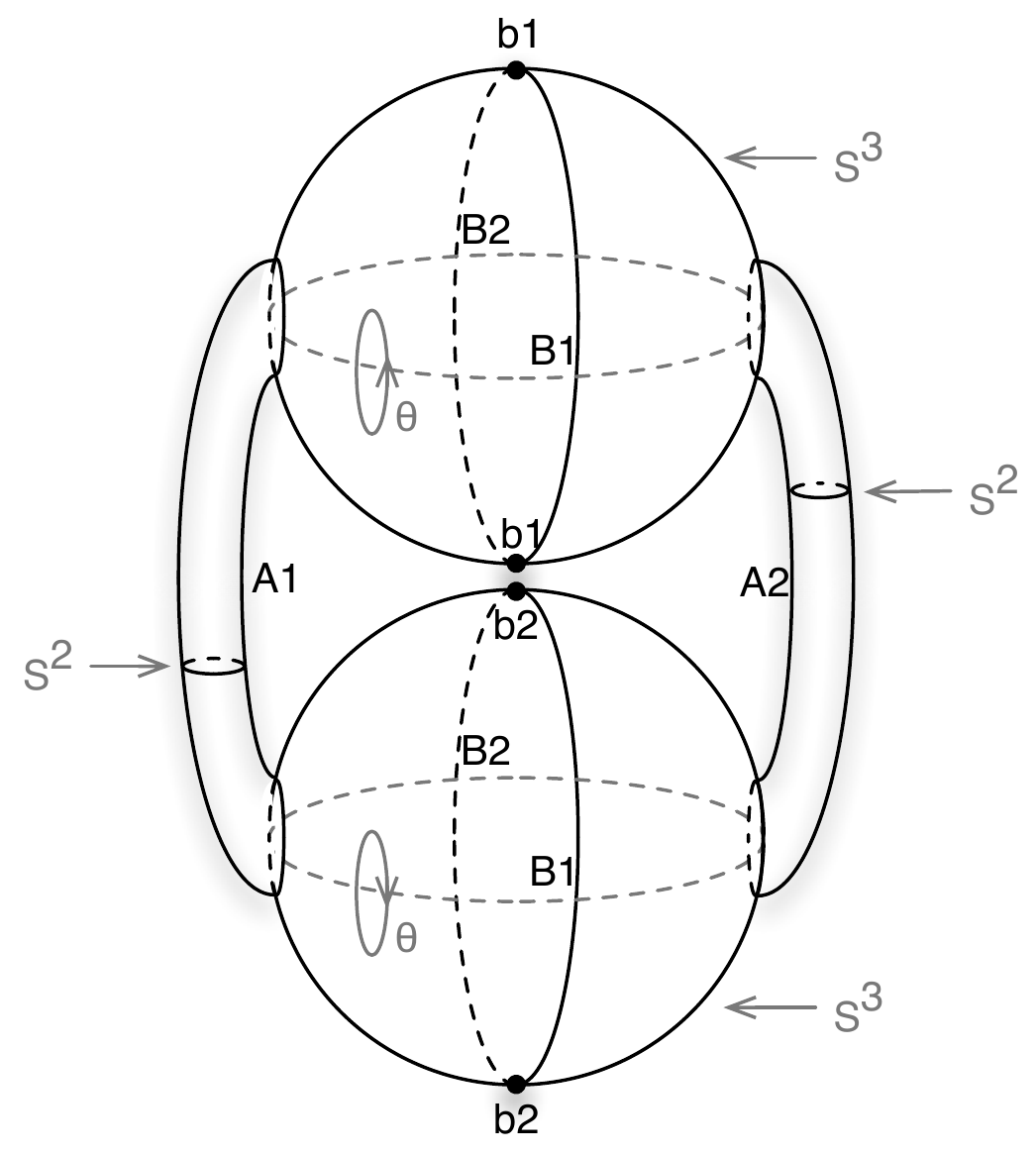}{6}{$S^2$ with two-component interface. For $n=2$, gluing two copies of Fig. \ref{fig:S2-2single} together gives a topology that can be thought of as two $S^3$'s that are joined along $n=2$ $S^2$'s. }
Thus we find
\beq
\frac{\textrm{tr}\rho^{n}_{A(S^{2},2)}}{\left(\textrm{tr}\rho_{A(S^{2},2)}\right)^n}=\frac{(Z(S^3))^{-n} Z(S^3)^2}{(Z(S^{3}))^{n}}=(Z(S^{3}))^{2(1-n)}=({{\cal S}_0}^0)^{2(1-n)}.
\eeq

It is not difficult to envision the generalization of this result to an $M$-component interface on $S^2$, and we find
\beq
S_A^{(M)} = M\ln {{\cal S}_0}^0.
\eeq

\subsection{$T^{2}$ with two-component AB interface}
\label{subsec:T2-2comp}

In the case of $T^{2}$ with a 2-component interface, there are a number of new choices to be made. There are essentially two distinct ways to slice the spatial surface, which we consider in turn.

\subsubsection{$T^2$: Disconnected B regions}
The first  possibility is shown in Fig. \ref{fig:T2-2single-1}. 
 \myfig{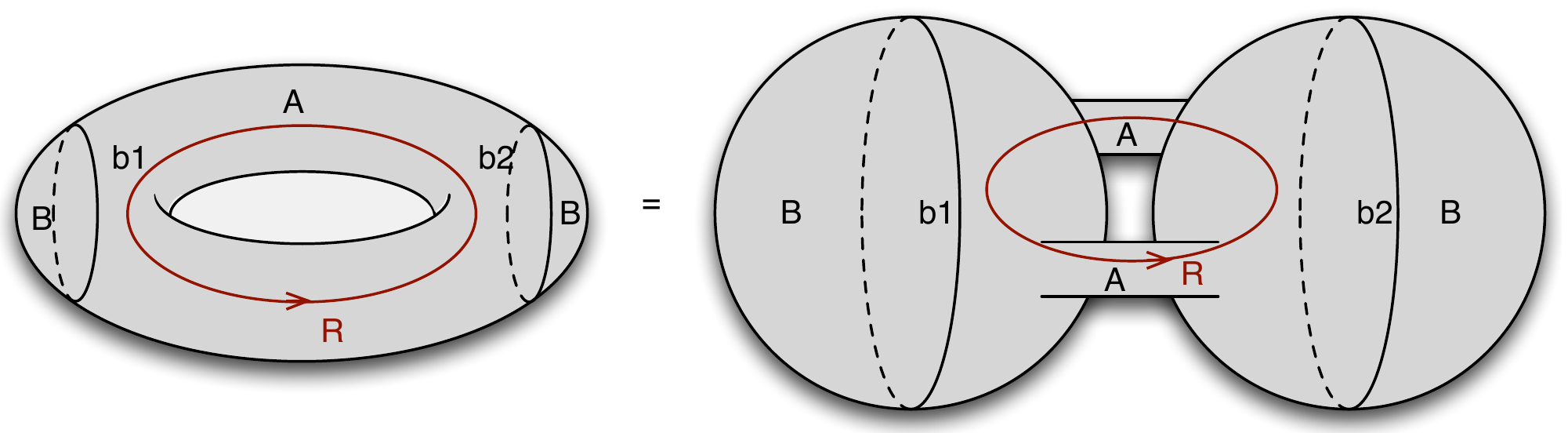}{9}
{The first of two ways of slicing the toroidal space with a two-component interface. B has two components. This may be thought of as two 3-balls joined by two tubes.}
Here, since we have learned that it is the features of the interface that matter to entanglement entropy, we expect that this would give the same result as in the last subsection. 
 \myfig{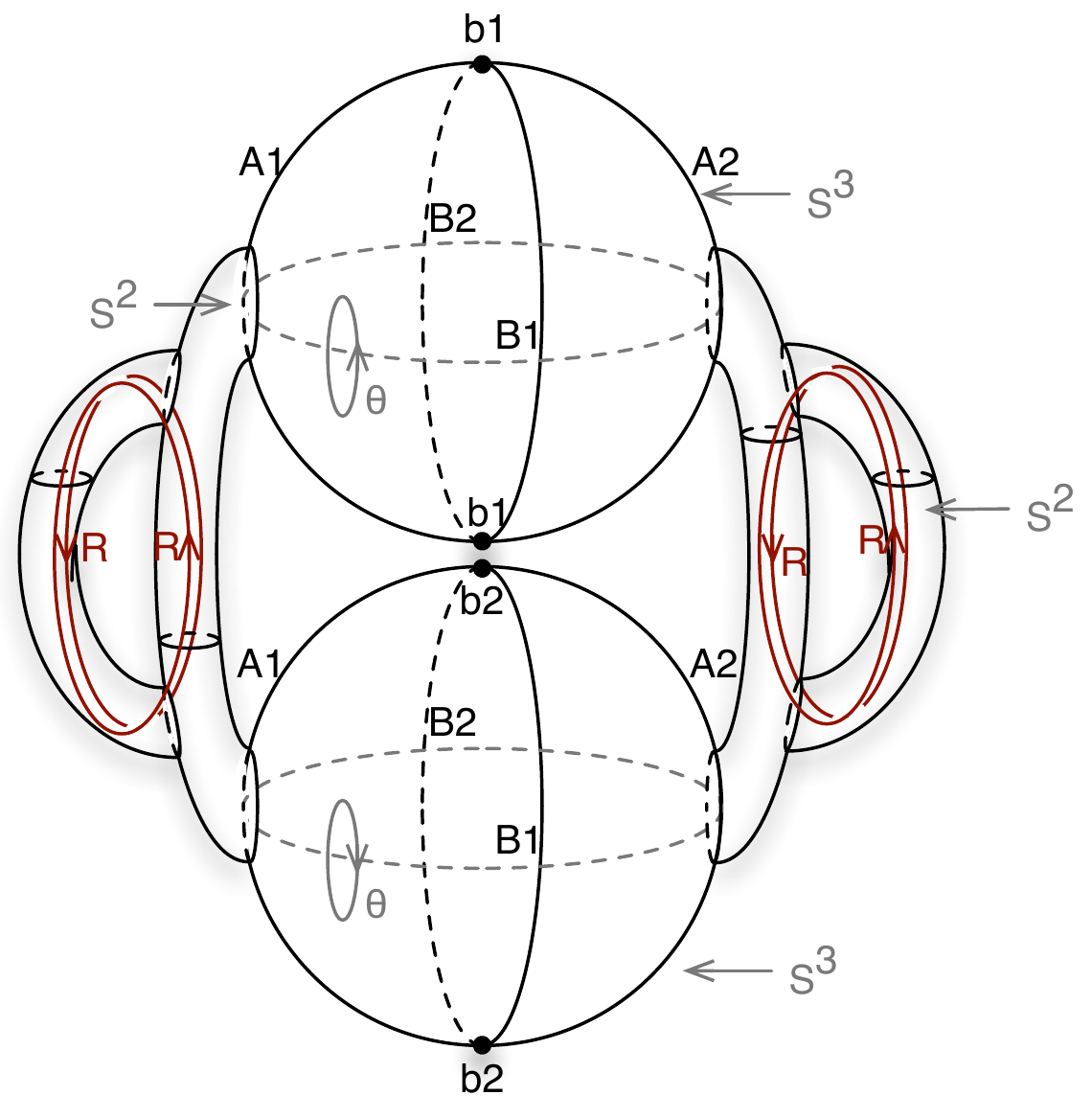}{6}
{The glued geometry for $n=2$ in the case of two disconnected B regions on a spatial torus.}
The glued geometry is shown in Fig. \ref{fig:T2-2double-1} for $n=2$. Note that the Wilson loops are located as shown in the figure. For this case, we find
\beq
\frac{\textrm{tr}\rho^{n}_{A(T^{2},2,\hat R)}}{\left(\textrm{tr}\rho_{A(T^{2},2,\hat R)}\right)^n}=
\frac{1}{(Z(S^{2}\times S^{1},\hat R,\hat{\overline R}))^{n}}
\frac{((Z(S^3))^{2}Z(S^{2}\times S^{1},\hat R,\hat{\overline R}))^{n}}{Z(S^3))^{2n}}=(Z(S^{3}))^{2(1-n)}=({{\cal S}_0}^0)^{2(1-n)},
\eeq
which indeed is the same result, for any representation $R$, as for the spherical topology with two interface components.

\subsubsection{$T^2$: Connected B region}
The second possibility, shown in Fig. \ref{fig:T2-2single-2}, will present new complications. The new feature here is that the Wilson loops thread through the interface between the A and B regions. We will find that this leads to a dependence on
the representation in the entanglement entropy.
 \myfig{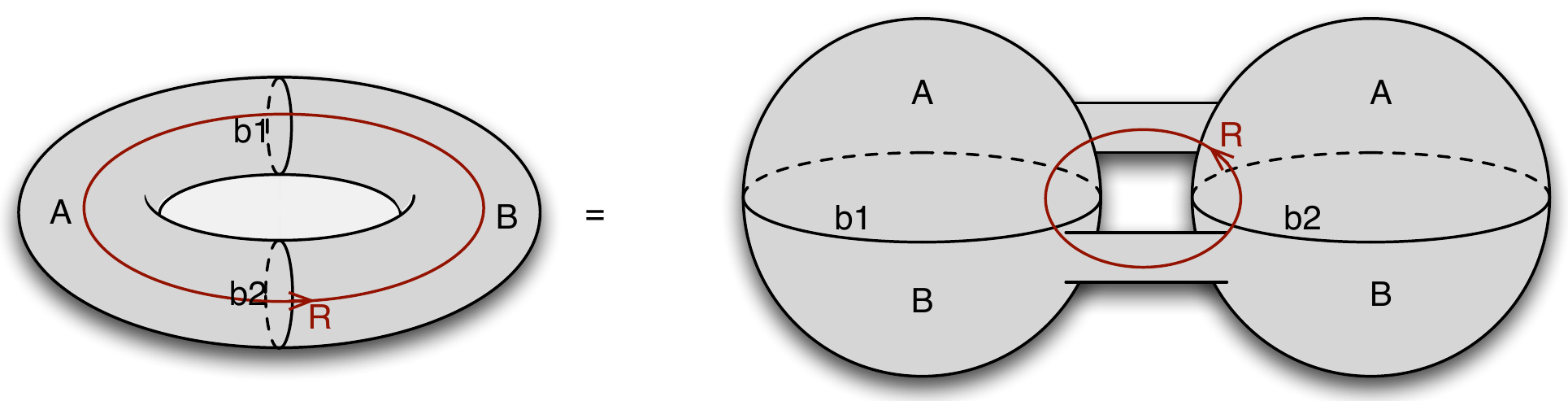}{9}
{The second of two ways of slicing the toroidal space with a two-component interface. B has a single component. Again, this may be thought of as two 3-balls joined by two tubes, but a Wilson loop threads the interface.}
Again, this computation can be thought of as a pair of 3-balls connected by tubes. Upon performing the gluing, we obtain a pair of $S^3$'s, connected along $2n$ $S^2$'s, with Wilson loops routed as shown in Fig. \ref{fig:T2-2double-2}.
 \myfig{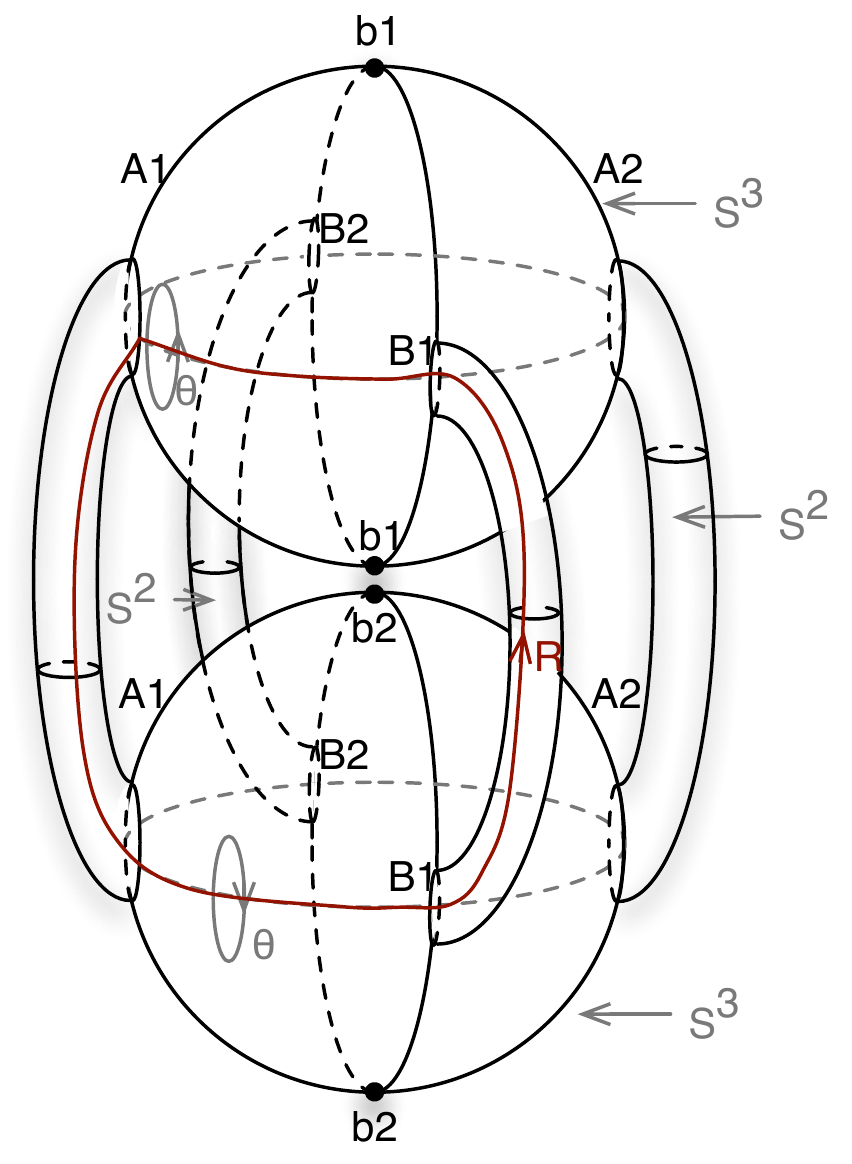}{5}{The glued geometry for $n=2$ in the case of a single connected B region on a spatial torus.}
 We find
\beq\label{T2gen}
\frac{\textrm{tr}(\rho^{n}_{A(T^{2},2,\hat{R}_j)})}{\textrm{tr}(\rho_{A(T^{2},2,\hat{R}_j)})^{n}}=\frac{Z(S^3,\hat{R}_j)^{-4n}Z(S^3,\hat{R}_j)^{2n}Z(S^3,\hat{R}_j)^{2}}{Z(S^{2}\times S^{1},\hat{R}_j,\hat{\overline{R}}_j)^{n}}=Z(S^{3},\hat{R}_j)^{2(1-n)}=({{\cal S}_0}^j)^{2(1-n)}.
\eeq
To obtain this result, we have generalized the formula \eqref{connsumZ} to the case where we also have a Wilson line. We can do this because the Hilbert space of Chern-Simons theory on $S^{2}$ with two complex conjugate charges  is one dimensional, just as $S^{2}$ with no punctures. As we explained earlier, the one-dimensionality allows us to cut and glue in half-$S^{3}$ caps; the same formula works for the half-$S^{3}$ with two conjugate punctures on the surface of the cap. The connection of these punctures inside the $S^{3}$ forms a Wilson loop.

Thus, in the numerator of Eq.\eqref{T2gen}, the first factor corrects for the inclusion of the endcaps (a total of 4 half-$S^3$'s per tube), the second factor comes from each of the $2n$ capped tubes (which are just $S^3$'s with a Wilson loop $\hat R_j$) while the last factor comes from each of the two ``large'' $S^3$'s that each have been capped $2n$ times. The routing of the Wilson loops through the original $2n$ tubes is such that after all of this surgery, there is a single Wilson loop in $\hat R_j$ on each of these large $S^3$'s.

It is then straightforward to show that for any state $|\psi\rangle=\sum_{j}\psi_{j}|{\hat{R}_j}\rangle$, 
\beqn
\frac{\textrm{tr}\rho^{n}_{A(T^{2},2,\psi)}}{\left(\textrm{tr}\rho_{A(T^{2},2,\psi)}\right)^n}&=&\frac{\textrm{tr}_{A}(\sum_{i,j}\psi_{j}\psi_{i}^{*}\textrm{tr}_{B}|{\hat{R}_j}\rangle\langle{\hat{R}_i}|)^{n}}{\left(\textrm{tr}\rho_{A(T^{2},2,\psi)}\right)^n}\nonumber\\
&=&\frac{\textrm{tr}_{A}(\sum_{j}\psi_{j}\psi_{j}^{*}\rho_{A(T^{2},2,\hat{R}_j)})^{n}}{\left(\textrm{tr}\rho_{A(T^{2},2,\psi)}\right)^n}\nonumber\\
&=&\frac{\prod_{p=1}^{n}\sum_{j_{p}}|\psi_{j_{p}}|^{2}\textrm{tr}(\rho_{A(T^{2},2,\hat{R}_{j_{1}})}\dots\rho_{A(T^{2},2,\hat{R}_{j_{n}})})}{\left(\textrm{tr}\rho_{A(T^{2},2,\psi)}\right)^n}\nonumber\\
&=&\frac{\sum_{j}|\psi_{j}|^{2n}\textrm{tr}(\rho_{A(T^{2},2,\hat{R}_{j})})^{n}}{\left(\textrm{tr}\rho_{A(T^{2},2,\psi)}\right)^n}\nonumber\\
&=&\frac{\sum_{j}|\psi_{j}|^{2n}({{\cal S}_0}^j)^{2(1-n)}}{(\sum_{j}|\psi_{j}|^{2})^{n}}.
\eeqn
Here we have used the fact that each $S^{2}$ we cut along should have total charge zero, otherwise the path integral vanishes. We thus obtain the entropy
\beq
S_{A(T^2,2,\psi)} = \sum_j \left[ 2|\psi_{j}|^{2}\ln {{\cal S}_0}^j - |\psi_j|^2\ln |\psi_j|^2\right].
\label{eq:40}
\eeq
Since we can interpret $|\psi_j|^2$ as a probability $p_j$, the second term has the familiar form $-p\ln p$. More precisely, note that this can be rewritten
\beqn
S_{A(T^2,2,\psi)} &=& 2\ln {{\cal S}_0}^0-\sum_j d_j^2 \left[ \frac{|\psi_j|^2}{d_j^2}\ln \frac{|\psi_j|^2}{d_j^2}\right].
\label{eq:addition1}
\eeqn

These calculations can be generalized to higher genus spatial surfaces, using similar techniques as we have displayed here. The entanglement entropy is sensitive to the topology only in cases where we choose carefully  the interface between the A and B regions. Finally, we note that Eq.\eqref{eq:addition1} is indicative of a more general result that says that the entanglement entropy depends on the number of interfaces, the states and how they fuse, and their quantum dimensions. Notice that in Eq.\eqref{eq:addition1}  the quantum dimension $d_j$ appears squared. This is so because there are two interface components in this case. In general there will be a factor of a quantum dimension for each interface component. However, in general the entanglement entropy will also depend on the non-universal  amplitudes in which the state appears in the conformal block. Hence in general the entanglement entropy depends on the universal properties of the topological field theory and on the specific form of the state.

\section{Quasiparticle Punctures}
\label{sec:punctures}

It is also of interest to consider entanglement entropy in the presence of quasi-particles. These will correspond to punctures on the spatial surface, and to each puncture we associate a representation $\hat R_j$. 
Here we will consider just  the simplest possibilities, but in so doing, we will explore how to write the entanglement entropy in a more convenient basis, that of conformal blocks.
\myfig{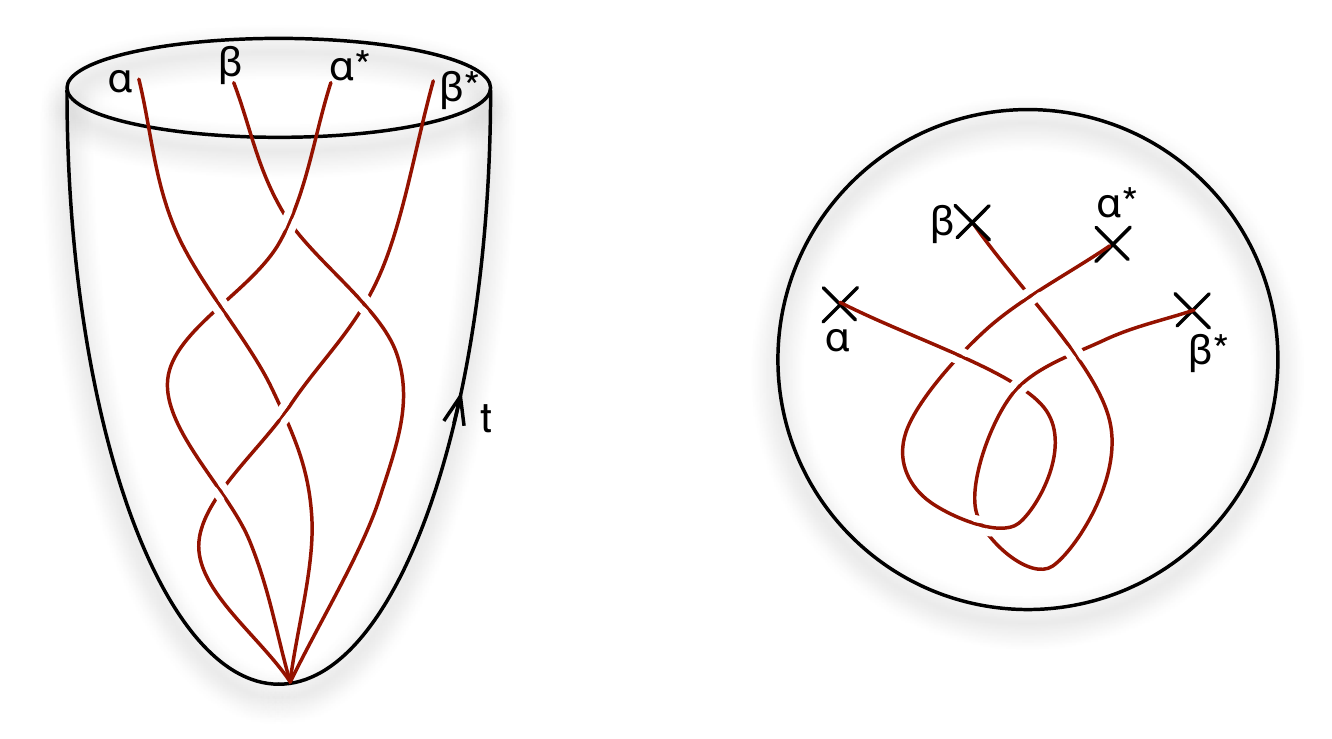}{8}{The wavefunctional with quasiparticles is related to a history with Wilson lines coming in from $t=-\infty$ to punctures on the spatial surface. On the left, time runs vertically to a spatial slice. On the right is the equivalent `radial time' view.}

\subsection{$S^{2}$ with four quasi-particles}
\label{subsec:S2-4}

A simple case is $S^{2}$ with four quasiparticles. We will focus on $\widehat{SU(N)}_k$, where $N\geq 2$ and $k\geq 2$, with punctures carrying 2 fundamental and 2 antifundamental representations on $S^2$. We will use $\hat\alpha$ and $\hat\alpha^*$ to denote fundamental and antifundamental representations, respectively. We may think of these punctures as being connected by oriented Wilson lines that extend into the interior; in this sense, they correspond to timelike Wilson lines extending in from $t=-\infty$. We note that these lines may braid. So we should expect that entanglement entropy may sense this braiding.
%
%
Now let us consider the entanglement entropy, where we simply divide the sphere into two halves. There are actually several distinct cases to consider. 
If the interface between A and B cuts one Wilson line (that is, A contains one puncture, say $\hat\alpha$, and B contains three, 
the result will be
\beq
S_{A}=\ln {{\cal S}_{0}}^{\hat\alpha}.
\eeq
If A and B each contain two punctures, there are two possibilities. The first possibility has $\hat\alpha$ and $\hat\alpha^*$ punctures in both A and B regions; in this case, the Wilson lines could connect  $\hat\alpha_A\hat\alpha^*_A$ and $\hat\alpha_B\hat\alpha^*_B$, or they could connect $\hat\alpha_A\hat\alpha^*_B$ and $\hat\alpha_B\hat\alpha^*_A$. The second possibility is that A contains two $\hat\alpha$'s and B two $\hat\alpha^*$'s; in this case, there are two possible connections of Wilson lines, and these differ by braiding. We will see that these choices correspond to choices of conformal blocks in the fusion of $\hat\alpha$ with $\hat\alpha^*$.

\subsubsection{B with $\alpha$ and $\alpha^*$}
\label{subsubsec:paired}

For $k\geq 2$, the Hilbert space on $S^2$ with 2 pairs of $\hat\alpha$ and $\hat\alpha^*$'s is two dimensional. Let's pick the two linearly independent states as follows.
 \myfig{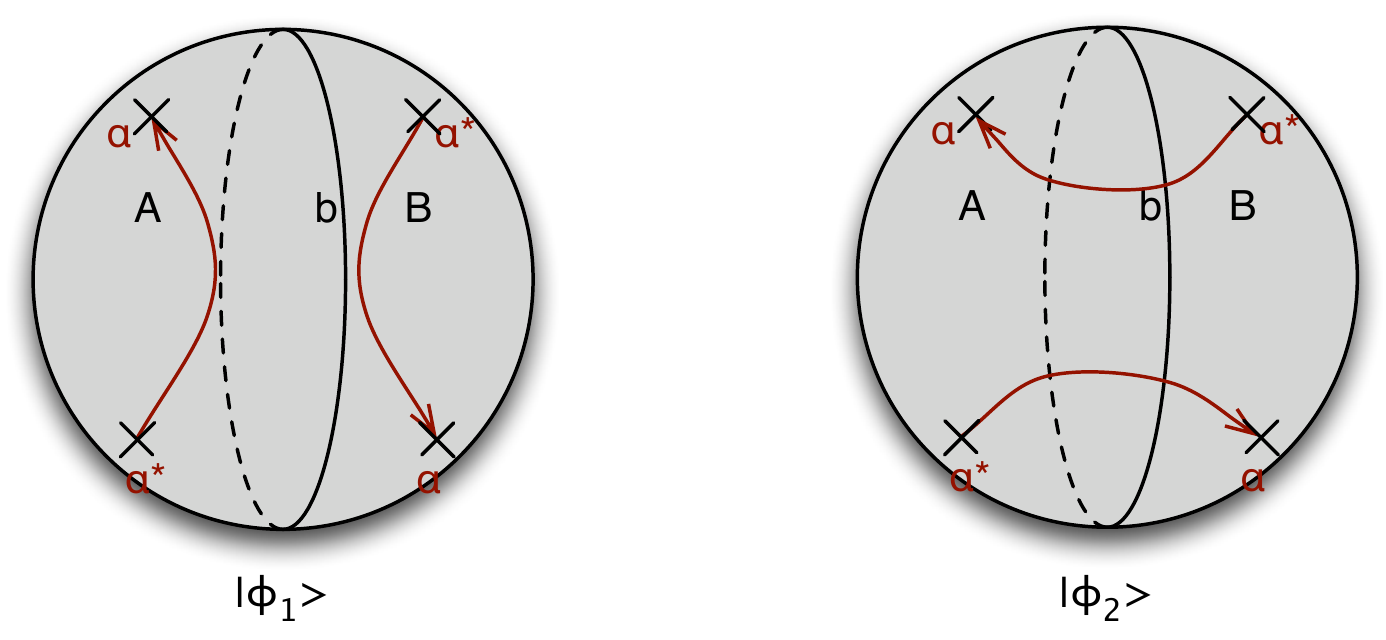}{8}
{The two states represented by Wilson lines connecting punctures.}
After gluing along B,\footnote{The gluing map is taken to identify punctures, and thus connects the Wilson lines.} there are four types of density matrix obtained; these are shown in Fig. \ref{fig:4p-pairedmatrices}.
 \myfig{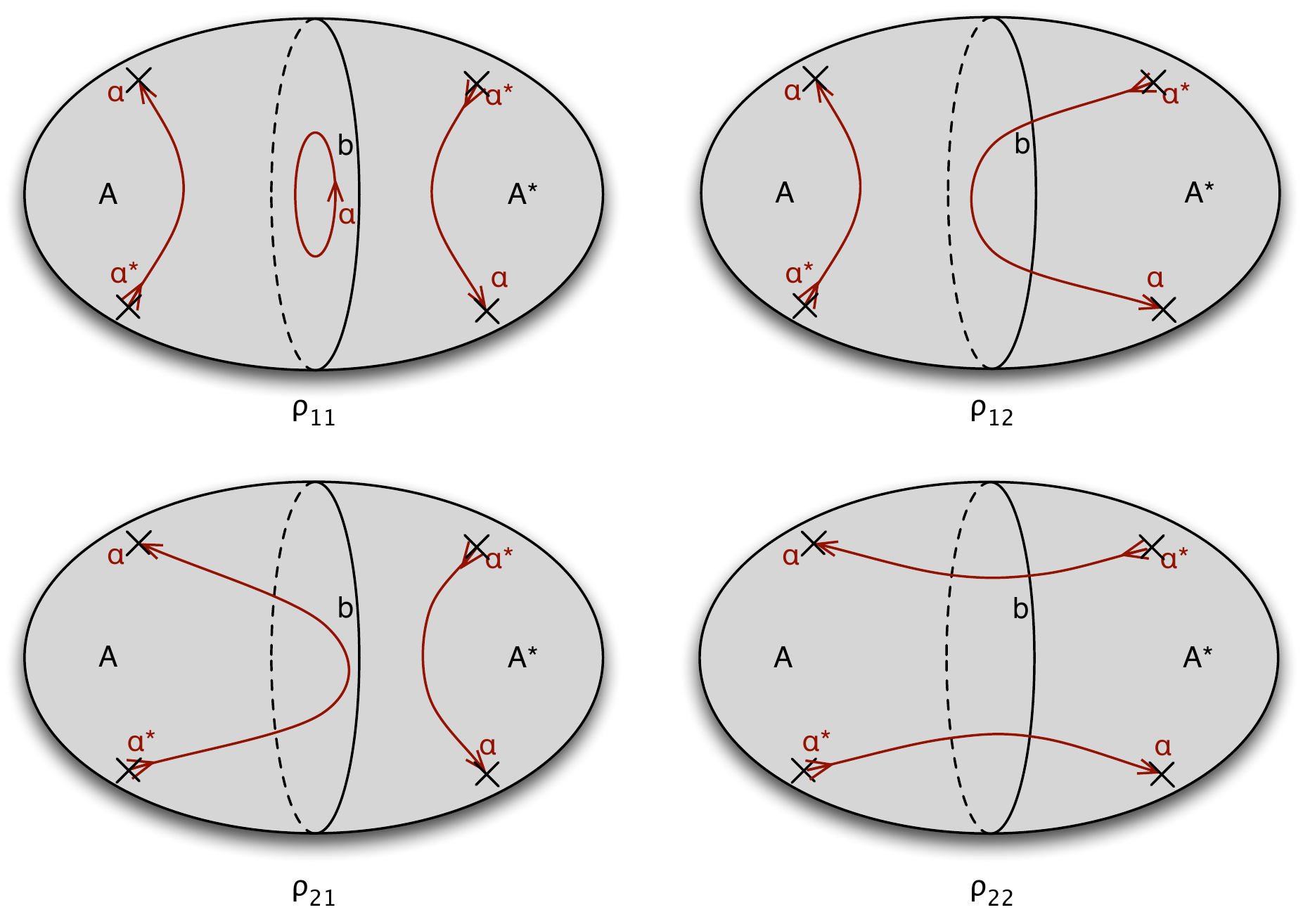}{9}
{Upon gluing to form $\rho_A$, we find four matrix elements.}
If we begin with a pure state $|\phi\rangle=a|\phi_1\rangle+b|\phi_2\rangle$, we have $\rho_A=aa^*\rho_{11}+ab^*\rho_{12}+a^*b\rho_{21}+bb^*\rho_{22}$. Gluing $n$ copies of these together to form $\textrm{tr}{\rho_{A}}^{n}$ gives rise to an $S^{3}$ made from  all the possible combinations of $\rho_{ij}$'s. To compute $\textrm{tr}{\rho_{A}}^{n}$, we need to identify the Wilson loops formed in each case. Each combination contains a number of fundamental Wilson loops, each of which contributes a factor $\frac{{{\cal S}_{0}}^{\hat\alpha}}{{{\cal S}_{0}}^{0}}=d_{\hat\alpha}$. For each appearance of $\rho_{11}$, there will be two such factors. For each factor of $\rho_{12}$ or $\rho_{21}$ there will be one such factor. Finally, factors of $\rho_{22}$ do not increase the number of loops; however, when the contribution to  $\textrm{tr}{\rho_{A}}^{n}$ is $n$ factors of $\rho_{22}$, there are two loops. Thus, we arrive at
\begin{eqnarray}
\frac{Z_{n}}{{{\cal S}_{0}}^{0}}&=&\sum_{j,k,l}\frac{n!}{j!k!l!(n-j-k-l)!}(aa^*)^{j}(ab^*)^{k}(a^*b)^{l}(bb^*)^{n-j-k-l}d_{\hat\alpha}^{2j+k+l}+(bb^{*})^{n}(d_{\hat\alpha}^{2}-1) \nonumber \\
&=&[aa^*d_{\hat\alpha}^{2}+(ab^*+a^*b)d_{\hat\alpha}+bb^*]^{n}+(bb^{*})^{n}(d_{\hat\alpha}^{2}-1).
\end{eqnarray}
After normalization,
\beqn
&&\frac{Z_{n}}{{Z_{1}}^{n}}=\\
&&({{\cal S}_{0}}^{0})^{1-n}
	\left\{
		\left[ \frac{aa^*d_{\hat\alpha}^{2}+(ab^*+a^*b)d_{\hat\alpha}+bb^*}
				{(aa^*+bb^*)d_{\hat\alpha}^{2}+(ab^*+a^*b)d_{\hat\alpha}}
		\right]^{n}
		+
		\left[\frac{bb^{*}}
			{(aa^*+bb^*)d_{\hat\alpha}^{2}+(ab^*+a^*b)d_{\hat\alpha}}
		\right]^{n}
		\left[ d_{\hat\alpha}^{2}-1
		\right]
	\right\}.\nonumber
\eeqn
Thus we find
\begin{eqnarray}
S_{A}&=&\ln{{\cal S}_{0}}^{0}-\lambda_{1}\ln\lambda_{1}-(d_{\hat\alpha}^{2}-1)\lambda_{2}\ln\lambda_{2}\label{4punctentent},
\end{eqnarray}
where
\begin{eqnarray}
\lambda_{1}&=&
\frac{|ad_{\hat\alpha}+b|^{2}}{|ad_{\hat\alpha}+b|^{2}+(d_{\hat\alpha}^{2}-1)|b|^{2}}, \nonumber \\
\lambda_{2}&=&
\frac{|b|^{2}}{|ad_{\hat\alpha}+b|^{2}+(d_{\hat\alpha}^{2}-1)|b|^{2}}.
\end{eqnarray}
We notice that while the definition of $|\phi_{1}\rangle$ and $|\phi_{2}\rangle$ makes the calculation transparent, they are not orthonormal. In fact, we have 
\begin{equation}
\langle\phi_{i}|\phi_{j}\rangle={{\cal S}_{0}}^{0}d_{\hat\alpha}\left(
\begin{array}{cc}
d_{\hat\alpha}&1\\
1&d_{\hat\alpha}
\end{array}
\right).
\end{equation}
If we define 
\begin{equation}
{|\phi'_{1}\rangle\choose|\phi'_{2}\rangle}=
\frac{1}{d_{\hat\alpha}\sqrt{{{\cal S}_{0}}^{0}}\sqrt{d_{\hat\alpha}^{2}-1}}\left(
\begin{array}{cc}
\sqrt{d_{\hat\alpha}^{2}-1}&0\\
-1&d_{\hat\alpha}
\end{array}
\right){|\phi_{1}\rangle\choose|\phi_{2}\rangle},
\end{equation}
we can show that the new states are orthonormal. In fact, the new states
$|\phi'_{1}\rangle$ and $|\phi'_{2}\rangle$ correspond to the conformal blocks associated with the trivial and adjoint representation $\hat\theta$, respectively, which appear in $\hat\alpha\times\hat\alpha^*$. We have just calculated the fusion matrix. $|\phi'_{1}\rangle$ and $|\phi'_{2}\rangle$ are the conformal blocks in one channel, while 
similarly $\frac{1}{d_{\hat\alpha}\sqrt{{{\cal S}_{0}}^{0}}}|\phi_{2}\rangle$ and $\frac{1}{d_{\hat\alpha}\sqrt{{{\cal S}_{0}}^{0}}\sqrt{d_{\hat\alpha}^{2}-1}}(-|\phi_{2}\rangle+d_{\hat\alpha}|\phi_{1}\rangle)$ should be the blocks in the other channel, using their relation, we can easily get the fusion matrix
\begin{equation}
F[{\tiny 
\begin{array}{cc}
\alpha&\alpha^{*}\\
\alpha^{*}&\alpha
\end{array}
}]=
\frac{1}{d_{\hat\alpha}}\left(
\begin{array}{cc}
1&\sqrt{(d_{\hat\alpha})^{2}-1}\\
\sqrt{(d_{\hat\alpha})^{2}-1}&-1
\end{array}
\right),
\end{equation}
where, $d_{\hat \alpha}$ is (as before) the quantum dimension of the fundamental representation $\hat \alpha$, and $\sqrt{d_{\hat \alpha}^2-1}\equiv d_{\hat \theta}$ is the quantum dimension of the adjoint representation $\hat \theta$.

In the conformal block basis the amplitudes become
\begin{equation}
{a'\choose b'}=\left(
\begin{array}{cc}
\sqrt{{{\cal S}_{0}}^{0}}d_{\hat\alpha}&\sqrt{{{\cal S}_{0}}^{0}}\\
0&\sqrt{{{\cal S}_{0}}^{0}}\sqrt{d_{\hat\alpha}^{2}-1}
\end{array}
\right){a\choose b}.
\end{equation}
In terms of the wavefunction in the orthonormal conformal block basis,
\beq
\lambda_{1}=\frac{|a'|^{2}}{|a'|^{2}+|b'|^{2}}, \quad\lambda_{2}=\frac{1}{(d_{\hat\alpha}^{2}-1)}\frac{|b'|^{2}}{|a'|^{2}+|b'|^{2}}.
\eeq
In the entropy formula Eq.\eqref{4punctentent}, we note that there is a degeneracy factor $(d_{\hat\alpha}^{2}-1)$ associated with the $|\phi'_{2}\rangle$ state.  Although in the example we have considered here, we took punctures carrying fundamentals, we see that the entanglement entropy is sensitive to the conformal block that the punctures in A (or B) may fuse to. This tells us if we label a state in terms of fusion or equivalently conformal blocks, we can read off the entanglement entropy directly from the representation around the interfaces.

\subsubsection{B with $\alpha^*$ and $\alpha^*$}
\label{subsubsec:not-paired}

To confirm this reasoning, let us compute carefully the second case, which consists of two fundamental punctures in A. In this case, there are two states, shown in Fig. \ref{fig:4p-npairedstates}.
 \myfig{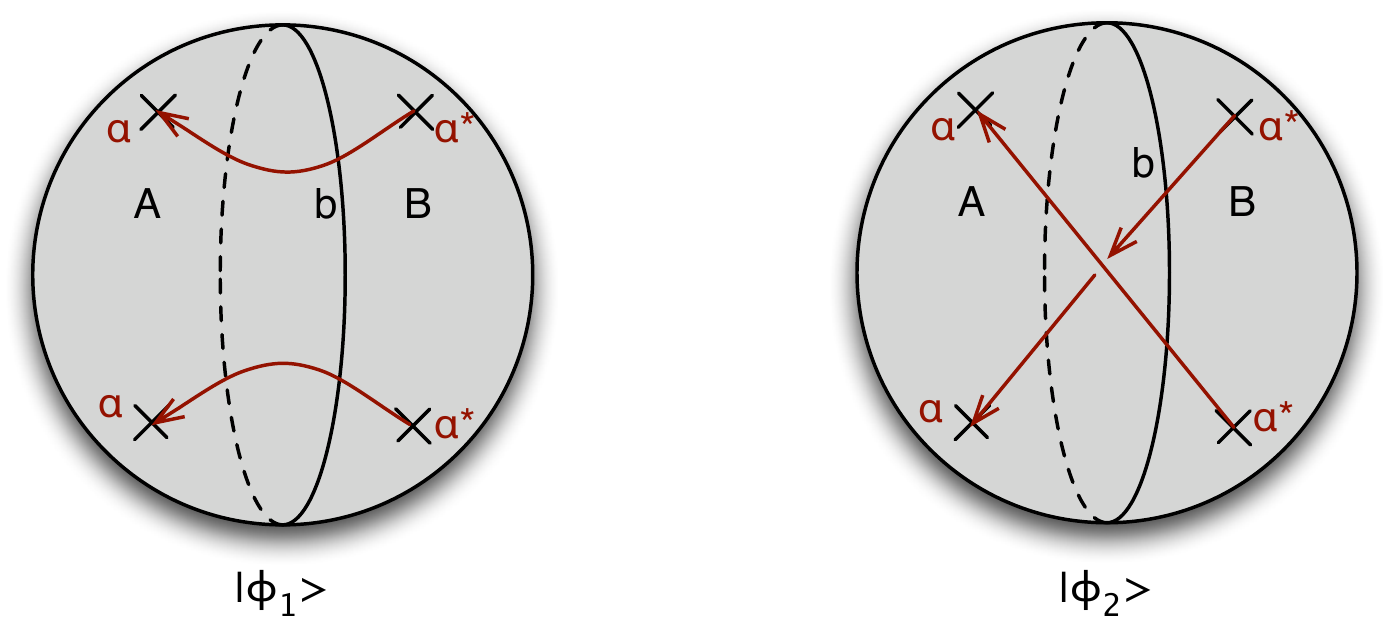}{8}
{The two states represented by Wilson lines connecting punctures.}
The fusion rules tell us that the representations cut by the interface are the symmetric and antisymmetric rank two tensor product of fundamental representations. Let's call them $\hat\sigma$ and $\hat\omega$, respectively. Of course in $SU(2)$, they are the same as $\hat\theta$ and $0$.
As before, we first glue along B and get four possible density matrices, as shown in Fig. \ref{fig:4p-npairedmatrices}. 
 \myfig{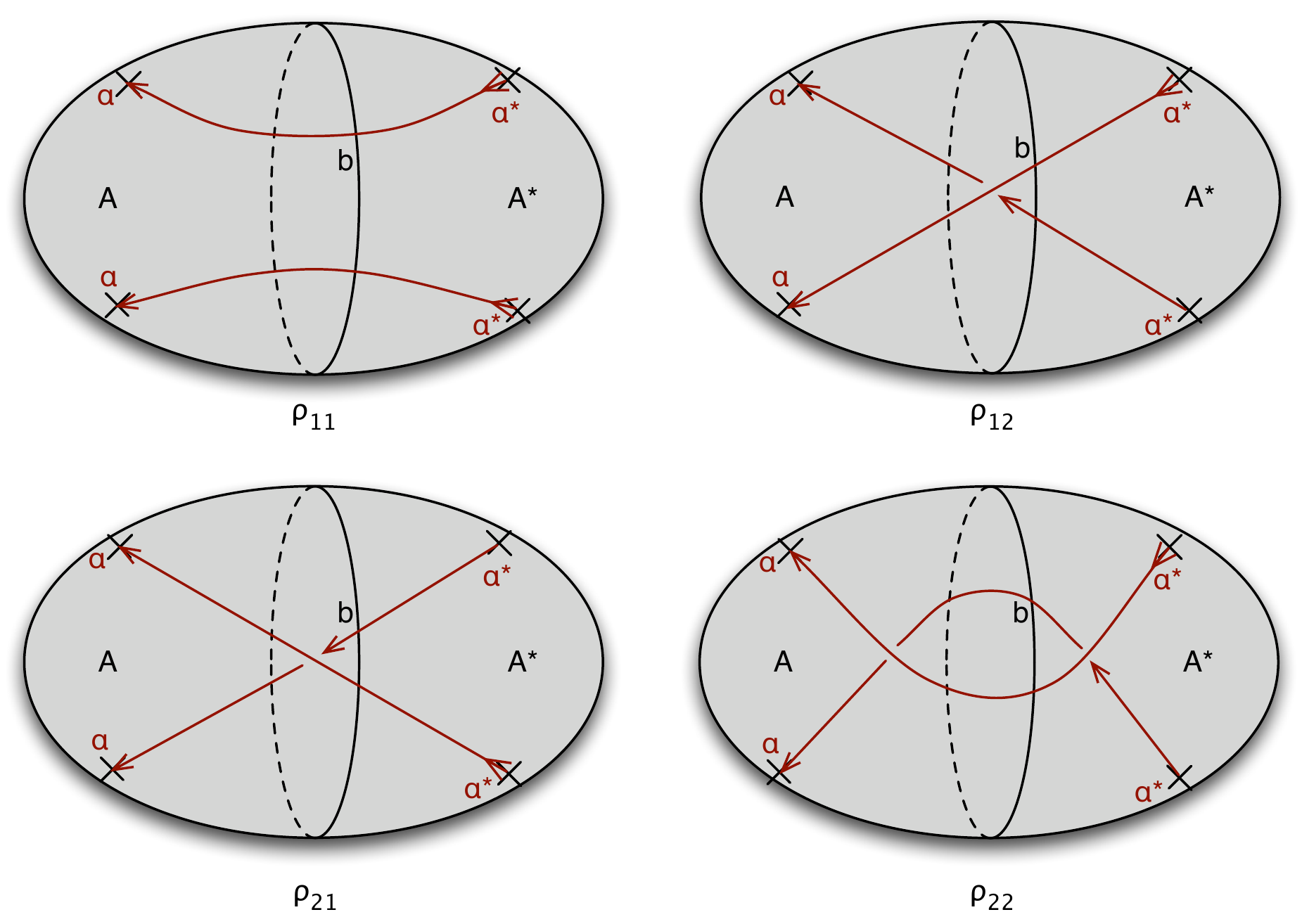}{9}
{Upon gluing to form $\rho_A$, we find four matrix elements. Note that $\rho_{12}$ and $\rho_{21}$ have the Wilson lines crossed in opposite senses.}
In both $\rho_{11}$ and $\rho_{22}$, we have two unlinked Wilson lines, so when they appear in $\textrm{tr}\rho_A^n$, they will not give rise to factors (as was the case for $\rho_{22}$ in the last example). 
Now, $\rho_{12}$ and $\rho_{21}$ have crossed Wilson lines with opposite orientation. In this case we have to be careful with the framing of the link and assign to each overcrossing a Dehn twist factor $t=e^{2\pi i h_{\hat \alpha}}$ (see Appendix B). So, if a $\rho_{12}$ appears together with $\rho_{21}$, they would just give an unlinked pair of lines, and thus the Dehn twists cancel each other. Thus only the difference in the number of each matters. To get the full $\textrm{tr}{\rho_{A}}^{n}$, we need to know the expectation value of a general braid with $j$ crossings; we call this $X_{j}$. We already know $X_{0}={{\cal S}_{0}}^{0}(d_{\hat\alpha})^{2}$ and $X_{1}={{\cal S}_{0}}^{0}td_{\hat\alpha}$. If we denote by $L_i$ a pair of lines with $i$ crossings, we have the skein relation $\alpha L_{+1}+\beta L_{0}+\gamma L_{-1}=0$, and then
\beq
\alpha X_{j}+\beta X_{j-1}+\gamma X_{j-2}=0.
\eeq
If we define $q=e^{-2\pi i /(N+k)}$, we have $\frac{\alpha}{\beta}=\frac{q^{-1/(2N)}}{q^{1/2}-q^{-1/2}}$ and $\frac{\gamma}{\beta}=-\frac{q^{1/(2N)}}{q^{1/2}-q^{-1/2}}$, thus
\beq
X_{j}+q^{\frac{1}{2}+\frac{1}{2N}}X_{j-1}=q^{-\frac{1}{2}+\frac{1}{2N}}(X_{j-1}+q^{\frac{1}{2}+\frac{1}{2N}}X_{j-2})=q^{(-\frac{1}{2}+\frac{1}{2N})(j-1)}{{\cal S}_{0}}^{0}[td_{\hat\alpha}+q^{\frac{1}{2}+\frac{1}{2N}}(d_{\hat\alpha})^{2}].
\eeq
Using the notation $[x]\equiv\frac{q^{x/2}-q^{-x/2}}{q^{1/2}-q^{-1/2}}$ (in which case $d_{\hat\alpha}=[N]$) and $t=q^{\frac{1-N^{2}}{2N}}$, we solve the difference equation to get
\beq
\frac{X_{j}}{{{\cal S}_{0}}^{0}}=(q^{\frac{1-N}{2N}})^{j}\frac{[N+1][N]}{[2]}
+(-q^{\frac{1+N}{2N}})^{j}\frac{[N][N-1]}{[2]}.
\eeq
This formula is valid for all integer $j$. Finally,
\begin{eqnarray}
\frac{Z_{n}}{{{\cal S}_{0}}^{0}}&=&\sum_{j,k,l}\frac{n!}{j!k!l!(n-j-k-l)!}(aa^*)^{j}(ab^*)^{k}(a^*b)^{l}(bb^*)^{n-j-k-l}\frac{X_{l-k}}{{{\cal S}_{0}}^{0}} \nonumber \\
&=&\frac{[N][N+1]}{[2]}|a+bq^{\frac{1-N}{2N}}|^{2n}+\frac{[N][N-1]}{[2]}|a-bq^{\frac{1+N}{2N}}|^{2n}.
\end{eqnarray}
We recognize in these expressions the quantum dimensions
\beq
d_{\hat\sigma}=\frac{[N][N+1]}{[2]},\ \ \ \ \  d_{\hat\omega}=\frac{[N][N-1]}{[2]}.
\eeq
The entropy then takes the form
\begin{eqnarray}
S_{A}&=&\ln{{\cal S}_{0}}^{0}-d_{\hat\omega}\lambda_{1}\ln\lambda_{1}-d_{\hat\sigma}\lambda_{2}\ln\lambda_{2},
\end{eqnarray}
where
\begin{eqnarray}
\lambda_{1}&=&\frac{|a-bq^{\frac{1+N}{2N}}|^{2}}{d_{\hat\sigma}|a+bq^{\frac{1-N}{2N}}|^{2}+d_{\hat\omega}|a-bq^{\frac{1+N}{2N}}|^{2}}, \nonumber \\
\lambda_{2}&=&\frac{|a+bq^{\frac{1-N}{2N}}|^{2}}{d_{\hat\sigma}|a+bq^{\frac{1-N}{2N}}|^{2}+d_{\hat\omega}|a-bq^{\frac{1+N}{2N}}|^{2}}.
\end{eqnarray}
$\lambda_{1}$ and $\lambda_{2}$ indicate an orthonormal basis, corresponding to the two conformal blocks. For the old basis, we have 
\beq
\langle\phi_{i}|\phi_{j}\rangle={{\cal S}_{0}}^{0}d_{\hat\alpha}\left(
\begin{array}{cc}
d_{\hat\alpha}&t\\
t^{*}&d_{\hat\alpha}
\end{array}
\right).
\eeq
We can define a new basis as follows,
\beq
{|\phi'_{1}\rangle\choose|\phi'_{2}\rangle}=
\frac{1}{[2]\sqrt{{{\cal S}_{0}}^{0}d_{\hat\sigma}d_{\hat\omega}}}\left(
\begin{array}{cc}
\sqrt{d_{\hat\sigma}}q^{-\frac{1}{2}}&-q^{-\frac{1}{2N}}\sqrt{d_{\hat\sigma}}\\
\sqrt{d_{\hat\omega}}q^{\frac{1}{2}}&q^{-\frac{1}{2N}}\sqrt{d_{\hat\omega}}
\end{array}
\right){|\phi_{1}\rangle\choose|\phi_{2}\rangle}.
\eeq
Again we can calculate the fusion matrix in this case. $|\phi'_{1}\rangle$ and $|\phi'_{2}\rangle$ are conformal blocks in the horizontal channel, while one choice of the conformal blocks in the vertical channel will be $\frac{1}{\sqrt{{{\cal S}_{0}}^{0}}d_{\hat\alpha}}|\phi_{1}\rangle$ and $\frac{q^{-\frac{N}{2}}}{\sqrt{{{\cal S}_{0}}^{0}}d_{\hat\alpha}\sqrt{d_{\hat\alpha}^{2}-1}}(|\phi_{1}\rangle-t^{*}d_{\hat\alpha}|\phi_{2}\rangle)$. Using their relation, we can calculate the fusion matrix for $\widehat{SU(N)}_k$  as  
\beq
F[{\tiny 
\begin{array}{cc}
\alpha&\alpha^{*}\\
\alpha&\alpha^{*}
\end{array}
}]=
\frac{1}{d_{\hat\alpha}}\left(
\begin{array}{cc}
\sqrt{d_{\hat\omega}}&\sqrt{d_{\hat\sigma}}\\
\sqrt{d_{\hat\sigma}}&-\sqrt{d_{\hat\omega}}
\end{array}
\right).
\eeq
When $N=2$ there's no difference between $\alpha$ and $\alpha^{*}$, and it matches the result of the previous subsection.

The wave function in the new conformal block basis is
\beq
{a'\choose b'}=\left(
\begin{array}{cc}
\sqrt{{{\cal S}_{0}}^{0}}\sqrt{d_{\hat\omega}}&-q^{\frac{1}{2N}+\frac{1}{2}}\sqrt{{{\cal S}_{0}}^{0}}\sqrt{d_{\hat\omega}}  \\
\sqrt{{{\cal S}_{0}}^{0}}\sqrt{d_{\hat\sigma}}&q^{\frac{1}{2N}-\frac{1}{2}}\sqrt{{{\cal S}_{0}}^{0}}\sqrt{d_{\hat\sigma}}
\end{array}
\right){a\choose b}.
\eeq
In terms of wave functions under the conformal block states,
\beq
\lambda_{1}=\frac{1}{d_{\hat\omega}}\frac{|a'|^{2}}{|a'|^{2}+|b'|^{2}}, \quad\lambda_{2}=\frac{1}{d_{\hat\sigma}}\frac{|b'|^{2}}{|a'|^{2}+|b'|^{2}}.
\eeq
giving the probability of finding a state in a given conformal block.

\subsection{$S^{2}$ with three quasiparticles}
\label{subsec:S2-3}

There are many other cases that we could consider; generically, they cannot be represented by ordinary Wilson lines. The simplest such case is the three-punctured sphere; in terms of Wilson lines attached to the punctures, this would look like a `string junction'. However, the density matrix $\rho_A$ can be thought of in terms of Wilson lines. And of course given what we have learned in the previous section, we know that the entanglement entropy can be computed directly in the conformal block basis. 

For example, let's put $\hat\alpha$, $\hat\alpha*$ and $\hat\theta$ on $S^{2}$ as in Fig. \ref{fig:3p-punctures}. 
 \myfig{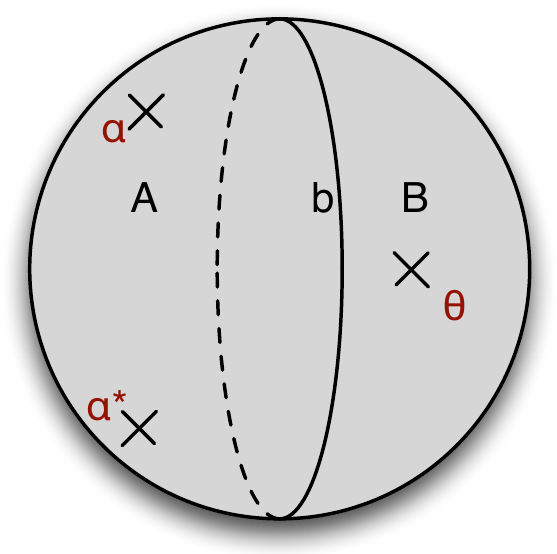}{4}
{Sphere with three punctures, chosen as representations $\alpha, \alpha^*, \theta$. }
Here the density matrix has the same conformal block as one of the states we found in the last section. Thus, up to a normalization factor, ${\rho_{\hat\theta}}_{A}=-\rho_{1}+d_{\hat\alpha}\rho_{2}$.
 \myfig{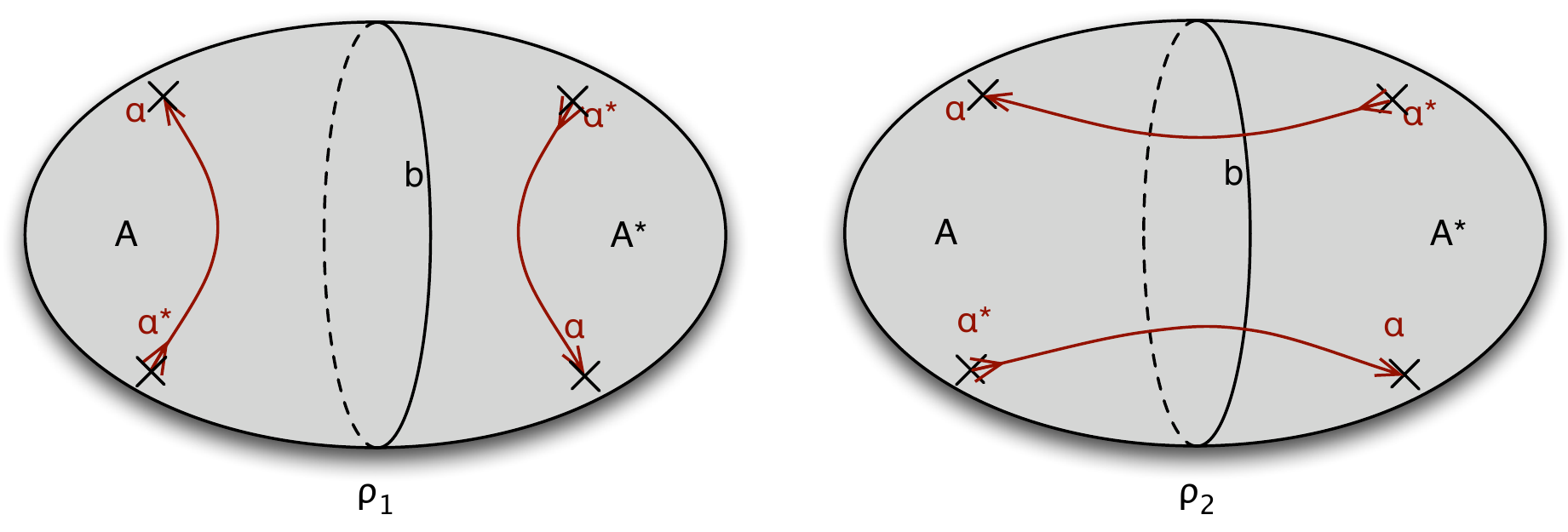}{9}
{The two density matrix elements in the case of the three-punctured sphere.}
Following the same construction as in the last section, we will get
\beq
\frac{Z_{n}}{{{\cal S}_{0}}^{0}}=(-d_{\hat\alpha}+d_{\hat\alpha})^{n}+(d_{\hat\alpha})^{n}d_{\hat\theta},\quad\frac{Z_{n}}{Z_{1}^{n}}=({{\cal S}_{0}}^{\hat\theta})^{1-n},\quad S_{A(\hat\alpha,\hat\alpha^*);\hat\theta}=\ln {{\cal S}_{0}}^{\hat\theta}.
\eeq
Clearly, there is a single conformal block contributing here.

Similarly for $\hat\alpha, \hat\alpha^{*}, \hat{0}$ insertions,  we find
\beq
S_{A(\hat\alpha,\hat\alpha^*);\hat 0}=\ln {{\cal S}_{0}}^{0}.
\eeq

Given these examples and further thought about the general case, we can generalize the three quasiparticle sphere to
\beq\label{entent3}
S_{A(i,j);k}=\ln {{\cal S}_0}^0 + \ln d_k.
\eeq

\subsection{Does the topological entropy depend on the entire $\mathcal{S}$-matrix?}
\label{sec:question}

In the previous discussion we have found that the entanglement entropy depends on the quantum dimensions which are determined by the top row of the $\mathcal{S}$-matrix. It is natural to ask if in other computations the other matrix elements of ${\cal S}$ will also enter. We will argue here that the answer is essentially negative. As an example, let us consider a case in which one might have expected that the other matrix elements matter, the case of a torus in a non-trivial state $i$ ({\it e.g.\/} a state created by a Wilson loop in representation ${\hat R}_i$) with a puncture in representation ${\hat R}_j$. One might anticipate that the entanglement entropy of the region represented in Fig. \ref{fig:T2-2single-2} would depend on ${\mathcal{S}_i}^j$. In fact it does not. The reason is that when computing $\textrm{tr} \rho_A^n$, any possible explicit dependence on ${\mathcal{S}_i}^j$ cancels out when properly normalizing $\rho_A$. As a consequence the result depends on the $\mathcal{S}$-matrix only implicitly through the fusion numbers ${N_{ij}}^k$. Given the structure of this result, it appears that this is a general property.

Another issue is the state dependence; in particular, we may have a situation in which the Hilbert spaces have dimensions greater than one, and hence there is at least implicit dependence on ${N_{ij}}^k$.\footnote{For $\widehat{SU(2)}_k$, the ${N_{ij}}^k$'s are either zero or one, and so this issue doesn't arise.}
A simple case of interest is four punctures on a sphere; suppose $i,j$ are in A while $k,\ell$ are in B and that both $i\times j$ and $k\times\ell$ contain a block $m$. If ${N_{ij}}^m$ and ${N_{k\ell}}^m$ are larger than one,\footnote{The explicit examples considered above in Section \ref{subsec:S2-4} have ${N_{ij}}^m\leq 1$.}  then $\rho_A$ is a matrix of rank $\min({N_{ij}}^m,{N_{k\ell}}^m)$. Each eigenvalue $p_\alpha$ of $\rho_A$ contributes a factor $-p_\alpha\ln (p_\alpha/d_m)$ to the entanglement entropy (in the case of a single component interface). This would be summed over the possible fusion channels $m$. In the case where there are multiple interface components, we can organize the calculation into fusions $A\to \{ m_j\}$ and $B\to\{m_j\}$, where $j=1,\ldots,I$ label the interface components. In this case, the rank of $\rho_A$ is $\min(N_A^{\{m_j\}},N_B^{\{m_j\}})$, with $N_A^{\{m_j\}}$ the fusion number of punctures in the $A$ region into the collection of blocks $\{m_j\}$. Each eigenvalue $p_\alpha$ of $\rho_A$ then contributes a factor $-p_\alpha\ln(p_\alpha/\prod_j d_{m_j})$, which should be summed over $\alpha$ and the fusion channels. In this sense, the entanglement entropy depends on the fusion rules, but in an implicit way. This may be generalized to higher genus. In such a case, we also keep track of the representation along each handle, and these can make a contribution to the fusion numbers $N$; apart from taking this into account, the entanglement entropy is computed as we have described here. All of the examples considered explicitly in this paper (all of which had $rank(\rho_A)=1$) may also be expressed in this language.

\section{Chern-Simons Theory of the Topological Entanglement Entropy of Fractional Quantum Hall States}
\label{sec:fqhe}

 The FQH states are topological fluids whose low energy effective field theory is a Chern-Simons gauge theory. As we saw in the preceding sections the entanglement properties of $SU(N)_k$ Chern-Simons gauge theories depend on the modular $\mathcal{S}$-matrix, which  yield the quantum dimensions, and on the fusion rules of the excitations. We also saw that the entanglement entropy depends on the topology of the surface and on the regions that are being observed, and that when the states are degenerate the entanglement properties naturally also depends on which state is considered.
  
  In this section we apply the general results we derived in the preceding sections for Chern-Simons gauge theories to the computation of the entanglement entropies for both Abelian and non-Abelian fractional quantum Hall states.   
The results we will derive here apply only in the strict topological limit, that is for systems in the thermodynamic limit and for observed regions of size $L$ much larger than any intrinsic length scale of the physical system. In doing so we can only obtain the universal topological entropies. It will suffice to identify which Chern-Simons describes each case of interest and to use the results of the preceding section to compute the entropies.

 The entanglement entropy  for the (Abelian) Laughlin FQH wave functions \cite{Laughlin83}, as well as for the non-Abelian FQH pfaffian wave functions \cite{moore-read-1991,Read-Rezayi-1999}, was calculated numerically recently in several papers\cite{haque07,zozulya07} 
 which attempted to extract the topological entropy $\gamma$ for these states. This is in practice difficult to do numerically due to the large non-topological area term which needs to be subtracted. Similarly, the computation of the topological entropy in the conceptually much simpler  $\mathbb{Z}_2$ topological phase of the quantum dimer model on a triangular lattice, which has a small but finite correlation length, presents similar difficulties\cite{furukawa-misguich-2007}. 
 The deconfined phases of $2+1$-dimensional discrete gauge theories are actually the simplest models of topological phases \cite{krauss89,preskill90,bais-1992}. For Kitaev's toric code state~\cite{kitaev-2003}, {\it i.e.\/} the ultra-deconfined limit of a $\mathbb{Z}_2$ gauge theory, a state with a vanishing correlation length, it is simple to compute the entropy \cite{hamma05,levin-wen-2006}. The (non-topological) effects of a finite correlation length in a topological phase have been discussed in detail recently \cite{papanikolaou-raman-fradkin-2007}. The scaling behavior of the entanglement entropy across a $\mathbb{Z}_2$ confinement-deconfinement phase transition was recently studied numerically \cite{hamma-2007}, as well as the role of thermal fluctuations on the behavior of the entropy in the $\mathbb{Z}_2$ topological state \cite{castelnovo07a}.

 To proceed we will need to identify the Chern-Simons theory appropriate for the FQH state of interest. There is a well developed body of theory which does that and it is reviewed in Appendix A. The identifications that we need are the following: 
 \begin{enumerate}
 \item
  For the Abelian (Laughlin) FQH states, at filling factor $\nu=1/m$ (with $m$ an odd integer) the effective field theory is an Abelian Chern-Simons gauge theory ${U(1)_m}$ (see Ref.\cite{moore-read-1991,wen-1995}). It is straightforward to extend these results to the case of general Abelian FQH states.
  \item
   The bosonic non-Abelian FQH states are described by a  Chern-Simons gauge theory for ${SU(2)_k}$, whereas the fermionic non-Abelian FQH states  are described by Chern-Simons gauge theories whose CFTs are cosets of the form $[\widehat{SU(2)/U(1)}]_2 \times \widehat{U(1)}$   (see Refs.\cite{moore-read-1991,fradkin-nayak-tsvelik-wilczek-1998, Read-Rezayi-1999,fradkin-nayak-schoutens-1999,ardonne-2002} and Appendix A).\footnote{This tensor product notation is ambiguous. The full RCFT  has an extended chiral algebra. Its primaries are those  in the tensor product which are local with respect to the current $J^+$ (defined below) with conformal dimension $1+M/2$.\cite{milovanovic-read-1996,Read-Rezayi-1999} In Section \ref{subsec:coset2} we describe these structures in detail.}
   \item
   The  Chern-Simons theory describing generalizations of the $p_x+ip_y$ superconductors 
   (see Ref.\cite{read-green-2000,stone-chung-2006,fendley-fisher-nayak-2007c}) have a coset CFT $\widehat{\left(SU(2)/U(1)\right)_k}$.
   \item
   The results presented here can be generalized to other non-Abelian FQH states of interest, {\it e.g.\/} the unpolarized non-Abelian states of Ref.\cite{ardonne-2007} (and references therein) which involve more complicated systems such as $SU(3)_2$ and others. We will not discuss these cases here.
\end{enumerate}

\subsection{${U(1)}_m$ Chern-Simons: The $\nu=\frac{1}{m}$ FQH Laughlin states}
\label{subsec:U1}

We will begin with a discussion of the $\nu=1/m$ FQH Laughlin states which correspond to a $U(1)$ Chern-Simons theory at level $m$. For the fermionic states $m$ is an odd integer, whereas for the bosonic states $m$ is an even integer. This case, and its connection with the modular $\mathcal{S}$-matrix and quantum dimensions, was discussed in great detail in Ref.\cite{fendley-fisher-nayak-2007c}. For completeness, here we present only a summary of the relevant results. 
The description of the edge states of the Laughlin states in terms of a compactified CFT is due to Wen.\cite{wen-1990,wen-1991}.

The $\widehat{U(1)}$ theory consists of a compact free chiral boson of compactification radius $R$. We will normalize\footnote{In string theory conventions, this corresponds to units $\alpha'=2$.} the field such that its correlator is $\langle\phi(z)\phi(0)\rangle\sim -\ln z$. There is a $U(1)$ current $J_{0}\sim i \partial\phi$ and operators ${\cal O}_Q \sim \exp(iQ\phi/R)$ of conformal dimension $h_Q=Q^2/2R^2$. If $Q\in\ZZ$, then ${\cal O}_Q$ is single-valued. 
The characters of this model are 
\beq
\chi_{n,w}(\tau)=\frac{q^{(n/R + wR/2)^2/2}}{\eta(q)},
\eeq
where $q=e^{2\pi i\tau}$.
In a rational CFT, we have that the radius is given by $R=\sqrt{2p'/p}$
where $p,p'$ are co-prime integers, in which case we can rewrite these characters as
\beq
\chi_{r,\ell}(\tau)=\frac{q^{pp'(r+\ell/2pp')^2}}{\eta(q)}.
\eeq
In this expression, $r\in\ZZ$ and $-pp'<\ell\leq pp'$. These can be organized into characters of an extended algebra generated by $J_{0}$ and operators $J_\pm$, a set which closes under the action of the modular group. Generally, we find
\beq\label{chioddppprime}
\chi_\ell(\tau)=\sum_{s\in\ZZ}\frac{q^{pp'(s+\ell/2pp')^2}}{\eta(\tau)},
\eeq
where $\ell\in -pp'+1,...,pp'$.\cite{DiFrancesco:1997nk} 
The modular ${\cal S}$-matrix for these characters is 
\beq
{{\cal S}_\ell}^{\ell'}=\frac{1}{\sqrt{2pp'}}e^{i\pi\ell\ell'/pp'},
\label{eq:Smatrix-abelian}
\eeq
 as can be easily established through Poisson resummation. We will use these formulae in later sections.

In the case where $pp'$ is even, this can be refined (that is, it is consistent (with respect to modular transformations) to consider a subsector of the Hilbert space) to 
\beq\label{chievenppprime}
\chi_{[n]}(\tau)\sim\chi_{\ell=2n}+\chi_{\ell=2n+pp'}=\sum_{s\in\ZZ}\frac{q^{m(s+n/m)^2/2}}{\eta(q)},
\eeq
where we have identified (when $pp'$ is even) $m=pp'/2$ and $n$ is in the range $0,1,...,m-1$. This is the case that obtains for the Abelian Laughlin states,\footnote{Specifically, we can take $p=2m$, $p'=1$, which gives radius $R=1/\sqrt{m}$. These values are of course ambiguous up to T-duality, which acts as $p\leftrightarrow p'$, $R\to 2/R$.} and we will refer to this theory as $\widehat{U(1)}_m$. The set of primaries are in one-to-one correspondence with the states of a bulk Chern-Simons theory at level $m$.
The extended current algebra is generated by $J$ and $J_\pm\sim exp(\pm i\sqrt{m}\ \phi)$, the latter having dimension $h_\pm=m/2$.  $J_+$ is the operator that shifts $n$ by $m$, leaving the character invariant and in the physical application, is interpreted as the electron. Requiring that primaries have local operator products with $J_\pm$, we find ${\cal O}_{2n/p}=\exp(in\phi/\sqrt{m})$, of dimensions $n^2/(2m)$; these correspond to the fractionally charged quasiparticles. 

Under a modular transformation, it is easy to establish (using Poisson resummation) that
\beq
\chi_{[n']}(-1/\tau)=\sum_\ell \frac{1}{\sqrt{m}}e^{2\pi i nn'/m}\chi_{[n]}(\tau).
\eeq
Thus, we read off the modular ${\cal S}$-matrix\footnote{The  $\mathcal{S}$-matrix of Eq.\eqref{eq:Smatrix-pfaffian} is complex, symmetric and unitary. It differs from the result of Ref.\cite{fendley-fisher-nayak-2007c} that found a real symmetric matrix. Nevertheless the quantum dimensions agree. }
\beq\label{suonek}
{{\cal S}_{[n']}}^{[n]} = \frac{1}{\sqrt{m}}e^{2\pi inn'/m},
\eeq
and thus the total quantum dimensions
 \beq
  {\cal D}=\left({{\cal S}_0}^0\right)^{-1}=\sqrt{\sum_\ell |d_\ell|^2 }=\sqrt{m}.
  \eeq
For the fermionic Laughlin states, $m$ is an odd integer so that $J_\pm$ is a 	fermionic operator, which usually identified with the electron. In the special case $m=3$ the electron has dimension $3/2$ and the theory is equivalent to a superconformal field theory. For the bosonic Laughlin states $m$ is even, and $J_\pm$ is a bosonic operator. In the special case $m=2$, the theory is equivalent to $\widehat{SU(2)}_1$.

\subsection{Coset $[\widehat{SU(2)/U(1)}]_k$ theories}
\label{subsec:coset1}

Chern-Simons theory whose CFT is the coset $[\widehat{SU(2)/U(1)}]_2$ describes the two-dimensional time reversal breaking superconductors with symmetry $p_x+ip_y$. In some sense this is the simplest system with non-Abelian statistics. Here we will consider the general case of the coset $[\widehat{SU(2)/U(1)}]_k$.

To construct this coset, we begin with the $\widehat{SU(2)}_k$ characters, which are labeled by $\ell = 2j =0,1,\ldots,k$, and decompose them with respect to $[\widehat{SU(2)/U(1)}]_k\times \widehat{U(1)}_k$
\beq
\chi_\ell^{SU(2)}(\tau)=\sum_{r=-k+1}^{k} \chi^{\mbox{\tiny coset}}_{\ell,r}(\tau)\chi^{U(1)}_r(\tau),
\eeq
where $\chi^{U(1)}_r(\tau)$ is given by Eq.\eqref{chioddppprime}, with $k=pp'$. 
Since we know the modular transformations of both the $SU(2)$ (Eq.\eqref{eq:Smatrix-SU2}) and $U(1)$ characters (Eq.\eqref{eq:Smatrix-abelian}), we read off
\beq
 \chi^{\mbox{\tiny coset}}_{\ell,r}(-1/\tau)= {{\cal S}_{(\ell,r)}}^{(\ell',r')} \chi^{\mbox{\tiny coset}}_{\ell',r'}(\tau),
\eeq 
with
\beq
{	{\cal S}_{(\ell,r)}	}^{(\ell',r')}=
{	({\cal S}^{SU(2)})_{\ell}	}^{\ell'}
{	({	{\cal S}^\dagger	}^{U(1)})_{r}	}^{r'}=
\sqrt{\frac{1}{k(k+2)}}		\sin\left[\frac{\pi(\ell+1)(\ell'+1)}{k+2}\right]
e^{-i\pi rr'/k}.
\eeq
This should be restricted to $r+\ell \equiv 0 \;  (\rm{mod}\ 2)$ (since there is such a correlation between $U(1)$ charges and $SU(2)$ representations).
Note though that in this formula, equivalent characters appear twice, as
$\chi^{\mbox{\tiny coset}}_{\ell;r}=\chi^{\mbox{\tiny coset}}_{k-\ell;r\pm k}$.
Accounting for these caveats, we find the independent characters, which lead to the final form of the ${\cal S}$-matrix:
\beq
{	{\cal S}_{(\ell,r)}	}^{(\ell',r')}=
\sqrt{\frac{4}{k(k+2)}}		\sin\left[\frac{\pi(\ell+1)(\ell'+1)}{k+2}\right]
e^{-i\pi rr'/k}.
\eeq
 For the case of most physical interest, we have $k=2$, and  the coset primaries may be taken to be $(0;0), (1;1)$ and $(0;2)$. This is in fact just the chiral Ising model with $(0;0)\sim I$, $(1;1)\sim\sigma$ and $(0;2)\sim\psi$. The ${\cal S}$-matrix for $\left(\widehat{SU(2)/U(1)}\right)_2 $ is
\beq
{\cal S}^{\mbox{\tiny coset}}_{k=2}=\frac{1}{2}\left(
\begin{array}{ccc}
1&\sqrt{2}&1\\
\sqrt{2}&0&-\sqrt{2}\\
1&-\sqrt{2}&1\\
\end{array}
\right),
\eeq
which agrees with the results of Ref.\cite{fendley-fisher-nayak-2007c}.

\subsection{ Moore-Read and Read-Rezayi FQH states: pfaffian and generalized parafermion states}
\label{subsec:coset2}

We now turn to the Moore-Read and Read-Rezayi non-Abelian FQH states, and their generalization. The filling factor of these states is $\nu=k/(Mk+2)$; $M$ even corresponds to bosonic states and $M$ odd to fermionic states \cite{Read-Rezayi-1999}. As discussed above (and in Appendix A), these states are described by $\widehat{[SU(2)/U(1)]_k} \times \widehat{U(1)}$ CFTs, with a suitably defined level for the $U(1)$.  Examples of these states are the well known Moore-Read pfaffian states. The the fermionic state with $k=2$ and $M=1$ has filling factor $1/2$ ($5/2$ in the experiment), and the related bosonic state at filling factor $\nu=1$ has $k=2$ and $M=0$. The states with $k>2$ are the Read-Rezayi parafermionic states. 

We will discuss both the general fermionic and bosonic states with fixed $k$ and $M$. The RCFT of interest is in all cases embedded in $\widehat{\left(SU(2)/U(1)\right)_k} \times \widehat{U(1)}_{k(Mk+2)}$. We will consider the cases of $k$ even and $k$ odd as their structure is somewhat different. Here we only present details for the simpler cases. The details of the derivations for the general case are given in Appendix C.

By reasoning similar to the above, the resulting ${\cal S}$-matrix can be obtained by multiplying coset and $U(1)$ characters. For the pfaffian state $k=2$, the coset is a $\mathbb{Z}_2$ parafermion. The resulting $\mathcal{S}$-matrix will, up to identifications, be given by
\beq
	{{\cal S}_{(\ell,r;s)}}^{(\ell',r';s')}=
        {({\cal S}^{\mbox{\tiny coset}}_{2})_{(\ell,r)}}^{(\ell',r')}
	{(\mathcal{S}^{U(1)_{4M+4}} )_s}^{s'}.
\eeq
Primaries of this theory will be given by products of the $\mathbb{Z}_2$ primaries $\{ I,\sigma,\psi\}$ with $U(1)_{4M+4}$ primaries of the form ${\cal O}_{\ell/p}$. We seek a set of such operators that close under operator products and are local with respect to a suitable extended current algebra, which will be generated by $J_\pm \sim\psi \; e^{\pm i\sqrt{M+1}\phi}$, where $\psi$ is the Majorana fermion of $\mathbb{Z}_2$. For simplicity, we will consider two cases here, $M=0$ (take $p=2$, $p'=1$, radius $R=1$) and $M=1$ (take $p=4$, $p'=1$, radius $R=\sqrt{1/2}$).

In the case of $k=2$ and $M=0$, we find the integer-weight $J_\pm \sim \psi \; e^{\pm i \phi}$ as suitable extended currents. Requiring locality of operator products, we then find that the primaries of this theory are given by $I, \psi,\sigma e^{i\phi/2}$ (all others are related to these by action of $J_\pm$). These in fact are just the primaries of $\widehat{SU(2)}_2$, as we should expect. This is the bosonic pfaffian state. The associated modular $\mathcal{S}$-matrix was given in Section II and Appendix C.

In the case of $k=2$ and $M=1$ (the fermionic pfaffian state), we find $J_\pm \sim \psi \; e^{\pm i\sqrt{2} \phi}$ as suitable extended currents. Requiring locality of operator products, we then find that the primaries of this theory are given by 
\beq\label{fpfstates}
I,\psi,\sigma e^{\pm i\phi/2\sqrt{2}},e^{\pm i\phi/\sqrt{2}}.
\eeq
This set closes under fusion (up to the action of $J_\pm$). These operators have weights\footnote{The notation $(\ell,r;s)$ represent the coset weights $(\ell,r)$ and the $U(1)$-charge $s$.} $(0,0;0)$, $(0,2;0)$, $(1,1;\pm1)$ and $(0,0;\pm2)$ respectively. We can then read off the ${\cal S}$-matrix:
\beq
{\cal S}=
\frac{1}{2\sqrt{2}}\left(
\begin{array}{cccccc}
1&1&\sqrt{2}&\sqrt{2}&1&1\\
1&1&-\sqrt{2}&-\sqrt{2}&1&1\\
\sqrt{2}&-\sqrt{2}&0&0&+i\sqrt{2}&-i\sqrt{2}\\
\sqrt{2}&-\sqrt{2}&0&0&-i\sqrt{2}&+i\sqrt{2}\\
1&1&i\sqrt{2}&-i\sqrt{2}&-1&-1\\
1&1&-i\sqrt{2}&+i\sqrt{2}&-1&-1
\end{array}
\right),
\label{eq:Smatrix-pfaffian}
\eeq
from which one can read-off the total quantum dimension is ${\cal D}=2\sqrt{2}$. This model can also be viewed as (the NS sector of) an $N=2$  superconformal current algebra, as the current $J_+$, having conformal weight $3/2$, can be viewed as a supercharge. So the operators listed in Eq.\eqref{fpfstates} are then viewed as superconformal primaries.\cite{Read-Rezayi-1999,milovanovic-read-1996}

We will now consider the interesting example of the parafermionic states at $k=3$ and $M=1$: the Read-Rezayi parafermionic state for fermions at filling factor $2+2/5$. The $k=3$ coset has primaries at
$(\ell,r) = (0,0),\ (1,\pm 1),\ (2,0),\ (3,\pm 1)$, 
which we will refer to as $I,\sigma_{\pm}, \epsilon, \psi_\pm$ respectively.
Explicitly, denoting $s_p\equiv \sin (\pi p/5)$, we have
\beqn
{\cal S}^{\mbox{\tiny coset}}_{k=3}
&=&\frac{2}{\sqrt{15}}\left(
\begin{array}{cccccc}
s_1&s_2&s_2&s_2&s_1&s_1\\
s_2&e^{-i\pi/3}s_1&e^{+i\pi/3}s_1&-s_1&-e^{-i\pi/3}s_2&-e^{+i\pi/3}s_2\\
s_2&e^{+i\pi/3}s_1&e^{-i\pi/3}s_1&-s_1&-e^{+i\pi/3}s_2&-e^{-i\pi/3}s_2\\
s_2&-s_1&-s_1&-s_1&s_{2}&s_{2}\\
s_1&-e^{-i\pi/3}s_2&-e^{+i\pi/3}s_2&s_{2}&-e^{-i\pi/3}s_1&-e^{+i\pi/3}s_1\\
s_1&-e^{+i\pi/3}s_2&-e^{-i\pi/3}s_2&s_{2}&-e^{+i\pi/3}s_1&-e^{-i\pi/3}s_1\\
\end{array}
\right).
\eeqn
For this case there is an extended algebra generated by the $h=3/2$ operator $Q_+=\psi_+ \; e^{5i\phi/\sqrt{15}}$, where $\phi$ is a free boson of the $U(1)$ theory that we are attaching \cite{milovanovic-read-1996}. Representative primaries are
$(\ell,r;s) = (0,0;0)$, $ (3,-1;1)$, $ (3,1;2)$, $ (0,0;3)$, $ (3,-1;4)$ and 
$(\ell,r;s) =  (2,0;0)$, $ (1,-1;1)$, $ (1,1;2)$, $ (2,0;3)$, $ (1,-1;4)$. 
One can check that these have local OPE's with $Q_+$ and are closed under fusion. As we will see, it is convenient to group them into groups of $k+2=5$, as given. The theory obtained this way is actually an $N=2$ superconformal theory, with supercharges $Q_\pm$ ($Q_-$ being $\psi_-\; e^{-5i\phi/\sqrt{15}}$). $Q_+$ groups collections of conformal primaries together, {\it i.e.\/}, 
$\{(0,0;0), (3,1;5), (3,-1;10)\}$, 
$\{(3,-1;1), (0,0;6), (3,1;11)\}$, 
$\{(3,1;2), (3,-1;7), (0,0;12)\}$,  
\\
$\{(0,0;3), (3,1;8), (3,-1;13)\}$,
and   
$\{(3,-1;4), (0,0;9), (3,1;14)\}$ 
and 
$\{(2,0;0), (1,1;5), (1,-1;10)\} $, 
\\
$\{(1,-1;1), (2,0;6), (1,1;11)\} $, ~~
$\{(1,1;2), (1,-1;7), (2,0;12)\} $,  
$\{(2,0;3), (1,1;8), (1,-1;13)\} $,  
and 
$\{(1,-1;4), (0,0;9), (1,1;14)\} $.
Each of these triplets represents a superconformal family.
When we compute the ${\cal S}$-matrix with respect to the extended symmetry, we treat these groupings as one. That is, computing the $\mathcal{S}$-matrix element on the grouping gives a $3\times 3$ identity matrix times a factor. We collect those factors into the following $\mathcal{S}$-matrix.
\beqn\label{k3M1}
{\cal S}^{FRR}_{k=3}&=&\frac{2}{5}\left(
\begin{array}{cc}
\sin (\pi /5)&\sin (2\pi /5)\\
\sin (2\pi /5)&-\sin (\pi /5)\\
\end{array}\right)
\otimes
\left(\begin{array}{ccccc}
1&1&1&1&1\\
1&\omega_2&\omega_4&\omega_1&\omega_3\\
1&\omega_4&\omega_3&\omega_2&\omega_1\\
1&\omega_1&\omega_2&\omega_3&\omega_4\\
1&\omega_3&\omega_1&\omega_4&\omega_2\\
\end{array}\right).
\eeqn
where we have used the $U(1)$ ${\cal S}$-matrix is ${{\cal S}_{s}}^{s'}=\frac{1}{\sqrt{15}}e^{2\pi i ss'/15}$.
Above we used the notation is $\omega_{p}=e^{2\pi i p/5}$. 
The coefficient out front is $\frac{2}{\sqrt{15}}\cdot \frac{1}{\sqrt{15}}\cdot \frac{15}{5}$, the factors being the coefficients of the coset $\mathcal{S}$-matrix, the $U(1)$ $\mathcal{S}$-matrix and the order of the automorphism (5 in 15), respectively.
Note that it is easy to read off then the total quantum dimension 
\beq
{\cal D}=\frac{1}{{{\cal S}_0}^0}=\frac{5}{2\sin(\pi/5)}=
{\cal D}=\sqrt{5 + 5(s_2/s_1)^2}=\sqrt{5 (1+\phi^2)},
\eeq
where here $\phi=(\sqrt{5}+1)/2$ denotes the Golden Ratio (not the chiral boson!).

In Appendix C it is shown that  for  general $k$ and $M$, the primaries are the highest weight states of the form
\beq
\psi_{(\ell,\ell-2[\frac{n}{M}])} \;
\exp \left(i\frac{\ell+nk-(Mk+2)[\frac{n}{M}]}{\sqrt{k(Mk+2)}}\phi\right) \quad \mbox{or}\quad
\psi_{(\ell,\ell-2[\frac{n-1}{M}])} \;
\exp \left(i\frac{\ell+nk-(Mk+2)[\frac{n-1}{M}]}{\sqrt{k(Mk+2)}}\phi \right),
\eeq
where $\psi_{(\ell,r)}$ are  $\mathbb{Z}_k$-parafermion primaries, $n$ and $\ell$ are integers (with a suitable range, see Appendix C), $[x]$ is the closest integer to $x$. For general $k$ the $\mathcal{S}$-matrix is given by
\beq
{\mathcal{S}_{\{\ell;n\}}}^{\{\ell';n'\}}=\frac{2}{\sqrt{(k+2)(Mk+2)}}\sin \left[\frac{\pi(\ell+1)(\ell'+1)}{k+2}\right] \; \exp \left(\frac{\pi i(-M\ell\ell'+2\ell n'+2\ell' n+2knn')}{Mk+2}\right).
\eeq
One can read off from this the total quantum dimension $\mathcal{D}$ for all $M$ and $k$, since
\beq
\frac{1}{\mathcal{D}}={{\cal S}_0}^0=\frac{2}{\sqrt{(k+2)(Mk+2)}} \; \sin\left(\frac{\pi}{k+2}\right).
\eeq

In Appendix C we also show that for general (odd) $k$, the modular $\mathcal{S}$-matrix may be put  into  the simpler form
\beq
{\mathcal{S}_{\{l;n\}}}^{\{l';n'\}}=\frac{2}{\sqrt{(k+2)(Mk+2)}} \sin\left[\frac{\pi(\ell+1)(\ell'+1)}{k+2}\right] (-1)^{\ell\ell'}\; 
\exp\left(\frac{2\pi i k nn'}{Mk+2}\right),
\eeq
with $\ell=0,\dots ,\frac{k+1}{2}$ and $n=0,\dots ,Mk+1$. The result of the $k=3$, $M=1$ example considered above matches with this.

\section{Conclusions}
\label{sec:conclusions}

In this paper we computed the entanglement entropy for Chern-Simons gauge theory with general gauge group. We considered the specific cases of $SU(N)_k$ and various cosets (of interest in the theory of FQH states). We have done this by direct computation in the $2+1$-dimensional gauge theory using surgery techniques. We found that the entanglement entropy for these theories can be written as a Chern-Simons path integral in a complicated 3-manifold whose details depend upon the topology of the spatial surface and the way it is partitioned to compute the entanglement. In all cases the entanglement entropy can be expressed in terms of modular features of the dual two-dimensional conformal field theory. We found that in general the entanglement entropy depends on the universal data of this topological field theory (that is the quantum dimensions, the fusion rules and the corresponding fusion numbers). However in cases in which, due to  topology the ground state is degenerate, the entanglement entropy may also depend on the choice of state.

\section*{Acknowledgments}
We thank Paul Fendley,  Eun-Ah Kim, Alexei Kitaev, Michael Levin, Joel Moore, Chetan Nayak, Stefanos Papanikolaou, John Preskill, and Kirill Shtengel  for many discussions. This work was supported in part by the National Science Foundation through the grant NSF DMR 0442537 (EF) and the US Department of Energy grant DE-FG02-91ER40709, and the Stanford Institute for Theoretical Physics (EF).

\appendix

\appsection{A}{Chern-Simons gauge theory and Fractional Quantum Hall States}
\label{appsection:CSFQHE}

In this appendix we review the connection between FQH syes and Chern-Simons thory. The FQH states are incompressible electron fluids which, due to the presence of the large magnetic field, have an explicitly broken time reversal invariance.  These {\em topological fluids} have a ground state degeneracy which depends only on the topology of the surface on which the fluids reside.\cite{wen-niu90} In their low energy and long distance (hydrodynamic) regime, the FQH fluids behave as topological fluids.  The excitations of these fluid states (``quasiparticles'') are vortices which, in general carry fractional charge and fractional (braid) statistics.\cite{arovas-1984} In other words, the long-distance correlations in this topological fluid are encoded in  the fractional charge of its vortices and, more significantly, in the non-local effects of fractional statistics. However, since these topological fluids are condensates of electrons, these quasiparticle (vortex) states are {\em local} with respect to states representing an electron. This condition, and the quantization of the electron charge $e$ play a key role in the properties of the effective theories of these fluids \cite{moore-read-1991}. This structure is also responsible for the extended symmetries in the corresponding CFTs that we described in the body of the paper.

 In the hydrodynamic regime {\it i.e.\/}, at energies low compared to the quasiparticle excitation energies and on length scales long compared with the magnetic length, the physical properties of these FQH fluids have been shown~\cite{zhang-hansson-kivelson-1989,read-1989,lopez-fradkin-1991,frohlich-zee-1991,wen-1995,wen-zee-1992,fradkin-nayak-tsvelik-wilczek-1998,fradkin-nayak-schoutens-1999} to be described by an effective (topological) quantum field theory, the Chern-Simons gauge theory in $2+1$ dimensions.\cite{witten89,deser82}

\subsection{The Abelian Quantum Hall States}
 \label{app:abelian}
 
 We will consider first the Laughlin states, whose wave functions for a system of $N$ particles at filling factor $\nu=1/m$ are \cite{Laughlin83}
 \begin{equation}
 \Psi_m(z_1,\ldots,z_N)=\prod_{i<j} \left(z_i-z_j\right)^m e^{\displaystyle{-\sum_{i=1}^N |z_i|^2/4\ell^2}}
 \label{eq:laughlin}
 \end{equation}
 where  $\{z_i\}$ are the complex coordinates of $N$ particles, and $\ell$ is the magnetic length. 
 Following Ref.\cite{wen-1995}, we write the effective field theory of  the Laughlin FQH states, which have filling fraction $\nu=1/m$ (with $m$ and odd integer for fermions and an even integer for bosons), as $U(1)_m$ Chern-Simons theory, whose Lagrangian density is
\begin{equation}
{\mathcal L}= \frac{m}{4\pi} \epsilon_{\mu \nu \lambda} {\mathcal A}_\mu \partial_\nu {\mathcal A}_\lambda
\label{eq:CSU(1)}
\end{equation}
up to irrelevant operators whose effects are negligible in this extreme infrared regime. The field strength of the gauge field ${\mathcal A}_\mu$ is essentially the hydrodynamic charge current of the topological fluid ${\mathcal J}_\mu$
\begin{equation}
{\mathcal J}_\mu=-\frac{e}{2\pi} \epsilon_{\mu \nu \lambda} \partial_\nu {\mathcal A}_\lambda
\label{eq:current}
\end{equation}
where $e$ is the electric charge. The (infinitely) massive quasiparticle bulk excitations, the vortices of the topological fluid, are represented in this limit by temporal Wilson loops (the world lines of these quasiparticles). On a manifold of genus $g$, the ground states of the FQH fluids are degenerate \cite{wen-niu90}. In the Chern-Simons description, the degeneracy is $m^g$, where $m$ is the (quantized) level of the Chern-Simons theory.

This hydrodynamic description generalizes to describe all other Abelian FQH states \cite{halperin-1983,jain-1989}. The corresponding effective field theory is a Chern-Simons gauge theory of a tensor product of $U(1)$ gauge groups at various levels, as well as  non-Abelian groups at level $1$ (which only have Abelian representations of the braid group) \cite{wen-zee-1992,lopez-fradkin-2001}. 
 Here we will consider only the simpler Laughlin states as the generalizations of our results to the other Abelian fluids is straightforward.

A direct consequence of the topological nature of the FQH fluids is that, for a physical system with a boundary, their excitations are gapless. These edge states are described by a chiral conformal field theory (CFT). \cite{wen-1991,wen-1990,wen-1992} In particular, there exists a one-to-one correspondence between the gapped bulk quasiparticles and the primary fields of the edge chiral CFT. At the edge, the effective CFT for a Laughlin state is a $U(1)_m$ chiral boson $\phi$ in $1+1$ dimensions.

The physical requirement that the quasiparticle edge states are local with respect to the electron operator leads to the compactification of this CFT, which now becomes a chiral rational conformal field theory (RCFT).\cite{moore-read-1991,wen-1995} Thus, the only operators allowed in the edge chiral RCFT must obey the condition of being invariant under the compactification condition  $ \phi \to \phi+2\pi R$, where $R$ is the compactification radius (see Section \ref{sec:fqhe}). The compactification condition leads to a truncation of the spectrum which has $m$ distinct sectors, the same as the ground state degeneracy on the torus. Since in the Abelian states all the states are one-dimensional representations of the braid group, it follows that all the quantum dimensions of these states are equal to unity, $d_i=1$, where here $i=1,\ldots,m$.

 \subsection{The non-Abelian Quantum Hall  States}
 \label{app:non-abelian}
 
 The non-Abelian FQH states are more interesting and have a more intricate structure. Although no simple unified effective field theory of all the non-Abelian states yet exists, in all the cases that have so far been studied the effective field theory contains at least a ${U(1)}_m$ gauge group at some level $m$ (for the charge sector), and a non-Abelian gauge group such as ${SU(2)}_k$ at some level $k$. For instance, for the Moore-Read pfaffian FQH states\cite{moore-read-1991}
 \begin{equation}
 \Psi_q(z_1,\ldots,z_N)= \textrm{Pf}\left(\frac{1}{z_i-z_j}\right) \; \prod_{i<j} \left(z_i-z_j\right)^m\; e^{\displaystyle{-\sum_{i=1}^N |z_i|^2/4\ell^2}}
 \label{eq:moore-read-m}
 \end{equation}
 where $\textrm{Pf}\left(\frac{1}{z_i-z_j}\right)$ is the pfaffian of the matrix. It has long been known \cite{moore-read-1991} that this wave function can be regarded as a correlator in an Euclidean two-dimensional CFT. Indeed, the Laughlin factor is simply the expectation value of a product of vertex operators of a chiral Euclidean boson $V_{\sqrt{m}}=e^{i \sqrt{m} \phi}$ (in a neutralizing background) and a correlator of majorana fermions $\psi$ in an Ising chiral Euclidean CFT. The theory of the edge states of the Moore-Read states is a chiral CFT described in Section \ref{sec:fqhe}.
 
 It turns out that for the {\em bosonic} state at $m=1$ (with filling factor $\nu=1$, which may hopefully be accessible to experiments in ultra-cold gases of bosons in rotating traps\cite{cooper-wilkin-gunn-2001}), the effective field theory in the bulk  is simply an $\widehat{SU(2)}_2$ Chern-Simons gauge theory \cite{fradkin-nayak-tsvelik-wilczek-1998,fradkin-nayak-schoutens-1999}, without any $U(1)$ factors. The excitation spectrum of this state consists of a particle with non-Abelian braid statistics, the Moore-Read ``non-abelion'', created by the Ising primary field $\sigma$, whose quantum dimension is $d_\sigma=\sqrt{2}$ (since the degeneracy of a state with $2n$ such vortices is $2^{n-1}$ \cite{nayak-wilczek-1996}), a Majorana fermion $\psi$ (with quantum dimension $d_\psi=1$), and the  identity field $I$ (the boson  also with quantum dimension $d_0=1$). 
 
 The two-dimensional time-reversal breaking superconducting state with symmetry $p_x+ip_y$, apparently observed in Sr$_2$RuO$_4$, is also closely related to paired-Hall pfaffian states \cite{greiter-wen-wilczek-1991}. It is also a non-Abelian state and it is also, up to an $U(1)$ factor, an $\widehat{SU(2)}_2$ state. Indeed, Fendley, Fisher and Nayak \cite{fendley-fisher-nayak-2007c} have computed the effective quantum dimension for this case as well and found that it is also equal to $\sqrt{4}$. 
 
 The {\em fermionic} pfaffian state at $m=2$ is the natural state
  to explain the observed plateau in the quantum Hall conductance at filling factor $\nu=5/2=2+1/2$. As can be seen from the structure of the Moore-Read states, {\it c.f.\/} Eq.\eqref{eq:moore-read-m}, the bosonic state at $m=1$ and the fermionic state at $m=2$ differ only by the Laughlin factor. Hence, one expects the fermionic state also to be connected to ${SU(2)}_2$ which, up to some caveats \cite{fradkin-nayak-tsvelik-wilczek-1998}, is essentially correct. The SU(2) symmetry of the bosonic case is dynamical and it is broken in the fermionic case due to the change in the compactification radius of the boson (see Refs. \cite{fradkin-nayak-tsvelik-wilczek-1998,fradkin-nayak-schoutens-1999,fradkin-huerta-zemba-2001}.) 
  Thus, one expects the the quantum dimensions of ${SU(2)}_2$ should play a role here too. However, the presence of the additional $U(1)$ factors, associated with the charge sector,  and the breaking of the $SU(2)$ symmetry changes the dimensions. Recently, Fendley, Fisher and Nayak \cite{fendley-fisher-nayak-2007c} have analyzed this case in detail. The upshot of their analysis is that the fermionic pfaffian state has a total of six primary fields: the identity $I$, two (conjugate) non-abelion primaries $\sigma e^{\pm i\phi/(2\sqrt{2})}$, the Majorana fermion $\psi$, and the Laughlin quasiparticle and quasihole $e^{\pm i \phi/\sqrt{2}}$. The quantum dimensions of these states are, respectively, $d_I=1$, $d_\sigma=\sqrt{2}$, $d_\psi=1$, and $1$ for the Laughlin vortices.

More interesting from the point of view of topological quantum computing, but not yet clearly seen in quantum Hall experiments, are the Read-Rezayi parafermionic states.\cite{Read-Rezayi-1999}  The Read-rezayi states are constructed in a manner analogous to that of the Moore-Read states. The main and important difference is that the pfaffian factor, which as we saw is equivalent to a correlator of Majorana fermions in a chiral Ising CFT, is replaced by a {\em parafermion} correlator of a parafermionic chiral CFT. In particular, the simplest bosonic parafermionic state can be represented in terms of the Chern-Simons theory ${SU(2)}_3$ \cite{fradkin-nayak-schoutens-1999}. The fermionic counterpart can also be understood in similar ways. The interest in this state stems from its non-Abelian vortex. As a consequence of their fusion rules, the topological degeneracies of these vortex states follow the Fibonacci sequence, and the quantum dimension of this non-Abelian vortex is the Golden Mean \cite{dassarma07}.

\appsection{B}{Calculation of the $SU(N)_k$ modular $\mathcal{S}$-matrix and Framing Factor}
\label{appsection:modularS}
\subsection{The $\mathcal{S}$-matrix Elements}
 \label{sec:smatrix}
 
We use (14.247) in \cite{DiFrancesco:1997nk} 
\begin{equation}
\frac{{{\cal S}_{\hat\delta}}^{\hat\lambda}}{{{\cal S}_{\hat\delta}}^{0}}=\gamma_{\hat\lambda}^{(\hat\delta)}=\chi_{\lambda}\left[\frac{-2\pi i(\delta+\rho)}{k+g}\right]
\end{equation}
to compute the $\mathcal{S}$-matrix elements. The notations used in this Appendix are also adopted from \cite{DiFrancesco:1997nk} and differ somewhat from the notations we used in the body of the paper. 

For $\widehat{SU(N)}_{k}$, there is a natural orthonormal  basis for the root lattice to compute the characters. It is constructed as follows. Pick $N$ dimensional unit lattice with unit vectors $\{\epsilon_{i}\}$, $i=1,\dots ,N$, then the simple roots of $SU(N)$ can be written as 
\begin{equation}
\alpha_{i}=\epsilon_{i}-\epsilon_{i+1}, \quad i=1,\dots ,N-1, 
\end{equation}
{\it i.e.\/} the root space sits in the $N-1$ dimensional subspace with $\sum_{i=1}^{N}n_{i}=0$ for any element $\sum_{i=1}^{N}n_{i}\epsilon_{i}$.  Any integral representation $\lambda=\sum_{i=1}^{N-1}\lambda_{i}\omega_{i}$, if expressed in the orthonormal basis, becomes \begin{equation}\lambda=\sum_{i=1}^{N}(l_{i}-\kappa)\epsilon_{i},\end{equation} where $l_{i}=\sum_{j=i}^{N-1}\lambda_{j}$ is the partition of the associate Young tableau, and $\kappa=\frac{1}{N}\sum_{j=1}^{N-1}j\lambda_{j}$. In particular, the Weyl vector becomes
\begin{equation}
\rho=\sum_{i=1}^{N-1}\omega_{i}=\sum_{i=1}^{N}(N-i-\frac{N-1}{2})\epsilon_{i}.
\end{equation}

Since
\begin{equation}
\chi_{\lambda}\left[\frac{-2\pi i(\delta+\rho)}{k+g}\right]=\frac{D_{\lambda+\rho}\left(\frac{-2\pi i(\delta+\rho)}{k+g}\right)}{D_{\rho}\left(\frac{-2\pi i(\delta+\rho)}{k+g}\right)}=\frac{\sum_{w\in W}\epsilon(w)e^{\left(w(\lambda+\rho),(\frac{-2\pi i(\delta+\rho)}{k+g})\right)}}{\sum_{w\in W}\epsilon(w)e^{\left(w\rho,(\frac{-2\pi i(\delta+\rho)}{k+g})\right)}},
\end{equation}
where Weyl group $W$ is simply the symmetric group of $\{\epsilon_{i}\}$, if we define $q=e^{\frac{-2\pi i}{k+g}}$, we will have 
\begin{eqnarray*}
\frac{{{\cal S}_{\hat\delta}}^{\hat\lambda}}{{{\cal S}_{\hat\delta}}^{0}}&=&\frac{\sum_{s\in {\cal S}_{N}}\epsilon(s) \prod_{i=1}^{N}q^{({l_{\lambda}}_{s_{i}}+N-s_{i}-\kappa_{\lambda}-\kappa_{\rho})({l_{\delta}}_{i}+N-i-\kappa_{\delta}-\kappa_{\rho})}}{\sum_{s\in {\cal S}_{N}}\epsilon(s) \prod_{i=1}^{N}q^{(N-s_{i}-\kappa_{\rho})({l_{\delta}}_{i}+N-i-\kappa_{\delta}-\kappa_{\rho})}}\\
&=&\frac{\textrm{det}[q^{({l_{\lambda}}_{i}+N-i-\kappa_{\lambda}-\kappa_{\rho})({l_{\delta}}_{j}+N-j-\kappa_{\delta}-\kappa_{\rho})}]}{\textrm{det}[q^{(N-i-\kappa_{\rho})({l_{\delta}}_{j}+N-j-\kappa_{\delta}-\kappa_{\rho})}]}\\
&=&q^{-\kappa_{\lambda}\sum_{j=1}^{N}({l_{\delta}}_{j}+N-j-\kappa_{\delta}-\kappa_{\rho})}\frac{\textrm{det}[q^{({l_{\lambda}}_{i}+N-i)({l_{\delta}}_{j}+N-j-\kappa_{\delta}-\kappa_{\rho})}]}{\textrm{det}[q^{(N-i)({l_{\delta}}_{j}+N-j-\kappa_{\delta}-\kappa_{\rho})}]}\\
&=&\frac{\textrm{det}[q^{({l_{\lambda}}_{i}+N-i)({l_{\delta}}_{j}+N-j-\kappa_{\delta}-\kappa_{\rho})}]}{\textrm{det}[q^{(N-i)({l_{\delta}}_{j}+N-j-\kappa_{\delta}-\kappa_{\rho})}]}\\
&=&q^{(-\kappa_{\delta}-\kappa_{\rho})\sum_{i=1}^{N}{l_{\lambda}}_{i}}\frac{\textrm{det}[q^{({l_{\lambda}}_{i}+N-i)({l_{\delta}}_{j}+N-j)}]}{\textrm{det}[q^{(N-i)({l_{\delta}}_{j}+N-j)}]}\\
&=&q^{-N(\kappa_{\delta}+\kappa_{\rho})\kappa_{\lambda}}{\cal S}_{\lambda}(\{q^{{l_{\delta}}_{j}+N-j}\}),
\end{eqnarray*}
where in the last step we have used the Schur function \begin{equation}{\cal S}_{\lambda}(\{x_{j}\})\equiv\frac{\textrm{det}[x_{j}^{{l_{\lambda}}_{i}+N-i}]}{\textrm{det}[x_{j}^{N-i}]}.\end{equation}
To calculate the Schur function specialized at $\{q^{{l_{\delta}}_{j}+N-j}\}$, we can use one of the \textit{Giambelli's Formula} (see Appendix A.1 around (A.5) of \cite{Fulton:1991} for detailed discussion), 
\begin{equation}{\cal S}_{\lambda}(\{x_{j}\})=\textrm{det}[E_{{\lambda^{T}}_{i}+j-i}],\end{equation}
where $\lambda^{T}$ is the transposed partition of $\lambda$, and $E_{k}$ are the elementary symmetric polynomials generated by 
\begin{equation}E(t)=\prod_{i=1}^{N}(1+x_{i}t)=\sum_{m=0}^{\infty}E_{m}t^{m},\end{equation}
and $E_{j}=0$ for $j<0$.

Now, in our case
\begin{equation}E(t)=\prod_{i=1}^{N}(1+q^{{l_{\delta}}_{j}+N-j}t),\end{equation}
and in principle by expanding $E(t)$ we can read off all the $E_{m}$ and calculate ${\cal S}_{\lambda}(\{q^{{l_{\delta}}_{j}+N-j}\})$. 
From (14.217) of \cite{DiFrancesco:1997nk} we can see that ${{\cal S}_{\hat\delta}}^{0}={{\cal S}_{0}}^{\hat\delta}$, thus
\begin{equation}{{\cal S}_{\hat\delta}}^{\hat\lambda}={{\cal S}_{0}}^{0}\frac{{{\cal S}_{\hat\delta}}^{\hat\lambda}}{{{\cal S}_{\hat\delta}}^{0}}\frac{{{\cal S}_{0}}^{\hat\delta}}{{{\cal S}_{0}}^{0}}.\end{equation}

To calculate ${{\cal S}_{0}}^{0}$, we need to use (14.217) of \cite{DiFrancesco:1997nk}. For $SU(N)_{k}$, $|\Delta_{+}|=\frac{N(N-1)}{2}$, $|P/Q\check\,|=N$, $g=N$, $r=N-1$, thus
\begin{eqnarray*}
{{\cal S}_{0}}^{0}
&=&i^{\frac{N(N-1)}{2}}\sqrt{\frac{1}{N (k+N)^{N-1}}}\textrm{det}[q^{(N-i-\kappa_{\rho})(N-j-\kappa_{\rho})}]\\
&=&i^{\frac{N(N-1)}{2}}\sqrt{\frac{1}{N (k+N)^{N-1}}}q^{-N(\kappa_{\rho})^{2}}\textrm{det}[q^{(N-i)(N-j)}]\\
&=&i^{\frac{N(N-1)}{2}}\sqrt{\frac{1}{N (k+N)^{N-1}}}q^{-N(\frac{N-1}{2})^{2}}\prod_{1\leq i<j\leq N}(q^{N-i}-q^{N-j})\\
&=&\sqrt{\frac{1}{N (k+N)^{N-1}}}\prod_{1\leq i<j\leq N}(2\sin\frac{\pi(j-i)}{k+N}).\\
\end{eqnarray*}

The quantum dimensions are as follows,
\begin{equation}d_{\hat\lambda}=\frac{{{\cal S}_{0}}^{\hat\lambda}}{{{\cal S}_{0}}^{0}}=q^{-N\kappa_{\rho}\kappa_{\lambda}}{\cal S}_{\lambda}(\{q^{N-j}\}).\end{equation}
First we calculate the generating function,
\begin{equation}E(t)=\prod_{i=1}^{N}(1+q^{N-i}t)=1+\sum_{m=1}^{N}t^{m}\prod_{r=1}^{m}\frac{q^{N}-q^{r-1}}{q^{r}-1},\end{equation}
and then plug into
\begin{equation}{\cal S}_{\lambda}(\{q^{N-j}\})=\textrm{det}[E_{{\lambda^{T}}_{i}+j-i}].\end{equation}
For example, for the fundamental representation $\alpha$ of $SU(N)$, the transpose $\alpha^{T}=\{1,0,0,\dots,0\}$, and
\begin{equation}{\cal S}_{\alpha}(\{q^{N-j}\})=\textrm{det}\left(
\begin{array}{ccccc}
E_{1}&E_{2}&\cdots&\cdots&E_{N}\\
0&E_{0}&E_{1}&\cdots&\\
0&0&E_{0}&\cdots&\\
\multicolumn{5}{c}{\dotfill}\\
&\cdots&\cdots&0&E_{0}
\end{array}
\right)=E_{1}=\frac{q^{N}-1}{q-1}.\end{equation}
Thus
\begin{equation}
d_{\hat\alpha}=q^{-N\frac{N-1}{2}\frac{1}{N}}\frac{q^{N}-1}{q-1}=\frac{q^{N/2}-q^{-N/2}}{q^{1/2}-q^{-1/2}}=[N].
\end{equation}
Here we have used the $q$-number notation, defined as $[x]=\frac{q^{x/2}-q^{-x/2}}{q^{1/2}-q^{-1/2}}$. Using this, we can write $E_{m}$ as 
\begin{equation}
E_{m}=\prod_{r=1}^{m}\frac{q^{N}-q^{r-1}}{q^{r}-1}=\prod_{r=1}^{m}q^{\frac{N-1}{2}}\frac{[N+1-r]}{[r]}=q^{\frac{(N-1)m}{2}}\prod_{r=1}^{m}\frac{[N+1-r]}{[r]}.
\end{equation}

We also want to check $d_{\hat\alpha*}$, for which $\alpha*^{T}=\{N-1,0,0,\dots,0,0\}$, and 
\begin{equation}{\cal S}_{\alpha*}(\{q^{N-j}\})=\textrm{det}\left(
\begin{array}{ccccc}
E_{N-1}&E_{N}&0&\cdots&0\\
0&E_{0}&E_{1}&\cdots&\\
0&0&E_{0}&\cdots&\\
\multicolumn{5}{c}{\dotfill}\\
&\cdots&\cdots&0&E_{0}
\end{array}
\right)=E_{N-1}=q^{\frac{(N-1)^{2}}{2}}[N].\end{equation}
Thus
\begin{equation}d_{\hat\alpha^*}=q^{-N\frac{N-1}{2}\frac{N-1}{N}+\frac{(N-1)^{2}}{2}}[N]=[N]=d_{\hat\alpha}.\end{equation}

For symmetric and antisymmetric rank two representations $\hat\sigma$ and $\hat\omega$ we have $\sigma^{T}=\{1,1,0,\dots,0,0\}$ and $\omega^{T}=\{2,0,0,\dots,0,0\}$, thus
\begin{equation}{\cal S}_{\sigma}(\{q^{N-j}\})=\textrm{det}\left(
\begin{array}{ccccc}
E_{1}&E_{2}&\cdots&\cdots&E_{N}\\
E_{0}&E_{1}&E_{2}&\cdots&\\
0&0&E_{0}&\cdots&\\
\multicolumn{5}{c}{\dotfill}\\
&\cdots&\cdots&0&E_{0}
\end{array}
\right)=E_{1}^{2}-E_{2}=q^{N-1}([N]^{2}-\frac{[N][N-1]}{[2]})=q^{N-1}\frac{[N][N+1]}{[2]},\end{equation}
and 
\begin{equation}{\cal S}_{\omega}(\{q^{N-j}\})=\textrm{det}\left(
\begin{array}{ccccc}
E_{2}&E_{3}&\cdots&\cdots&E_{N}\\
0&E_{0}&E_{1}&\cdots&\\
0&0&E_{0}&\cdots&\\
\multicolumn{5}{c}{\dotfill}\\
&\cdots&\cdots&0&E_{0}
\end{array}
\right)=E_{2}=q^{N-1}\frac{[N][N-1]}{[2]}.\end{equation}
Thus
\begin{equation}d_{\hat\sigma}=q^{-N\frac{N-1}{2}\frac{2}{N}+(N-1)}\frac{[N][N+1]}{[2]}=\frac{[N][N+1]}{[2]},\end{equation}
and
\begin{equation}d_{\hat\omega}=q^{-N\frac{N-1}{2}\frac{2}{N}+(N-1)}\frac{[N][N-1]}{[2]}=\frac{[N][N-1]}{[2]}.\end{equation}
In the weak coupling limit when $k\to\infty$, $[x]\to x$, the quantum dimensions we just calculated match their classical values.

The last thing we want to calculate is ${{\cal S}_{\hat\alpha}}^{\hat\alpha}$. Following the procedure described before, the partitions $\alpha=\alpha^{T}=\{1,0,0,\dots,0\}$, so 
\begin{equation}{\cal S}_{\alpha}(\{q^{{l_{\alpha}}_{j}+N-j}\})=E_{1}(\{q^{{l_{\alpha}}_{j}+N-j}\}),\end{equation}
where $E_{1}(\{q^{{l_{\alpha}}_{j}+N-j}\})=\sum_{j=1}^{N}q^{{l_{\alpha}}_{j}+N-j}=q^{N}+\frac{q^{N-1}-1}{q-1}=q^{N}+q^{\frac{N-2}{2}}[N-1]$. Thus
\begin{equation}\frac{{{\cal S}_{\hat\alpha}}^{\hat\alpha}}{{{\cal S}_{0}}^{0}}=\frac{{{\cal S}_{\hat\alpha}}^{\hat\alpha}}{{{\cal S}_{\hat\alpha}}^{0}}\frac{{{\cal S}_{0}}^{\hat\alpha}}{{{\cal S}_{0}}^{0}}=q^{-(1+\frac{N(N-1)}{2})\frac{1}{N}}(q^{N}+q^{\frac{N-2}{2}}[N-1])[N]=q^{-\frac{1}{N}-\frac{1}{2}}(q^{\frac{N+2}{2}}+[N-1])[N].\end{equation}

 \subsection{The Framing Factor and the Skein Relation}
 \label{sec:framing}
 
For any representation $\hat\lambda$, the unit of Dehn twist factor is $t=e^{2\pi i h_{\hat\lambda}}$, where
\begin{equation}h_{\hat\lambda}=\frac{(\lambda,\lambda+2\rho)}{2(k+g)}.\end{equation}
For the fundamental representation $\hat\alpha$ of $\widehat{SU(N)}_{k}$, 
\begin{equation}h_{\hat\alpha}=\frac{(\omega_{1},\omega_{1}+2\sum_{i}\omega_{i})}{2(k+g)}=\frac{F_{11}+2\sum_{i}F_{1i}}{2(k+N)}=\frac{N^{2}-1}{2N(k+N)},\end{equation}
where $F$ here is the quadratic form matrix of $SU(N)$, which is the inverse of the Cartan matrix. and 
\begin{equation}t=q^{\frac{1-N^{2}}{2N}}.\end{equation}
In order to solve $\alpha L_{+1}+\beta L_{0}+\gamma L_{-1}=0$, we complete the path integral in two different ways and get
\begin{equation}\frac{\alpha}{\beta} t d_{\hat\alpha}+ (d_{\hat\alpha})^{2}+\frac{\gamma}{\beta}t^{*}d_{\hat\alpha}=0, \end{equation}
and
\begin{equation}\frac{\alpha}{\beta}(d_{\hat\alpha})^{2} + t^{*}d_{\hat\alpha}+\frac{\gamma}{\beta}\frac{{{\cal S}_{\hat\alpha}}^{\hat\alpha}}{{{\cal S}_{0}}^{0}}=0. \end{equation}
Notice unlike in \cite{witten89}, we keep the framing factor explicitly here. 
\begin{equation}\frac{\alpha}{\beta}=t^{*}\frac{1-t^{2}\frac{{{\cal S}_{\hat\alpha}}^{\hat\alpha}}{{{\cal S}_{0}}^{0}}}{t^{2}\frac{{{\cal S}_{\hat\alpha}}^{\hat\alpha}}{{{\cal S}_{\hat\alpha}}^{0}}-d_{\hat\alpha}}=t^{*}\frac{1-[N]^{2}+q^{-\frac{N}{2}}[N-1][N+1][N](q^{1/2}-q^{-1/2})}{-q^{-\frac{N}{2}}[N-1][N+1](q^{1/2}-q^{-1/2})}=\frac{q^{-\frac{1}{2N}}}{q^{1/2}-q^{-1/2}},\end{equation}
\begin{equation}\frac{\gamma}{\beta}=t\frac{(d_{\hat\alpha})^{2}-1}{t^{2}\frac{{{\cal S}_{\hat\alpha}}^{\hat\alpha}}{{{\cal S}_{\hat\alpha}}^{0}}-d_{\hat\alpha}}=t\frac{[N]^{2}-1}{-q^{-\frac{N}{2}}[N-1][N+1](q^{1/2}-q^{-1/2})}=-\frac{q^{\frac{1}{2N}}}{q^{1/2}-q^{-1/2}}=(\frac{\alpha}{\beta})^{*}.\end{equation}

\appsection{C}{$\cal{S}$-matrix of the $\left(\widehat{SU(2)/U(1)}\right)_{k}\times\widehat{U(1)}_{k(Mk+2)}$ RCFT}
\label{sec:parafermion}

In this appendix we will calculate the primaries and $\cal{S}$-matrix of the $\left(\widehat{SU(2)/U(1)}\right)_{k}\times\widehat{U(1)}_{k(Mk+2)}$ theory. For simplicity we will assume $k$ is odd for the moment, and discuss  even $k$ later on. The fields in the $\mathbb{Z}_{k}$-parafermion CFT, (the coset $\left(\widehat{SU(2)/U(1)}\right)_{k}$), are labeled by the $SU(2)$ charge and its $U(1)$ subgroup charge $(\ell,r)$, where we take them as twice the traditional values, so that they are integers. For $SU(2)_k$, $\ell=0,\dots, k$ and $\ell-r\equiv 0\,\bmod{2}$. We also have the identification
\begin{equation}(\ell,r)\equiv (k-\ell,r\pm k)\equiv (\ell,r\pm 2k)\dots.\end{equation}
Using this identification we can always map $(\ell,r)$ for $\ell>\frac{k+2}{2}$ to $(k-\ell,r\pm k)$. So we can restrict to $0\leq \ell\leq\frac{k+1}{2}$. We will use $\psi_{(\ell,r)}$ for the corresponding fields and $\chi_{(\ell,r)}$ their characters in the coset theory.

With these properties, one can see that the states form $\mathbb{Z}_{k}$-loops generated by $\psi$ or $\psi^{\dagger}$, where $\psi$ and $\psi^{\dagger}$ are the parafermion fields that appeared in the original $\widehat{SU(2)}_{k}$ currents $J_+\sim\psi e^{i\phi\sqrt{2/k}}$ and $J_-\sim\psi^{\dagger} e^{-i\phi\sqrt{2/k}}$. 
We can identify $\psi\sim\psi_{(0,2)}$ and $\psi^{\dagger}\sim\psi_{(0,-2)}$ for which $h_{\psi}=h_{\psi^{\dagger}}=1-1/k$. The other fields in the $\widehat{SU(2)}_{k}$, $\psi_{(\ell,r)}e^{ir\phi/\sqrt{2k}}$, have $U(1)$ charge $r$. The whole multiplet under  the $\widehat{SU(2)}_{k}$ current algebra has character 
\begin{equation}
\frac{1}{\eta(\tau)}\sum_{t=-\infty}^{\infty}\chi_{(\ell,\ell-2t)}q^{(\ell-2t)^{2}/4k}=\frac{1}{\eta(\tau)}\sum_{t=0}^{k-1}\chi_{(\ell,\ell-2t)}\sum_{p=-\infty}^{\infty}q^{(\ell-2t+2pk)^{2}/4k}.
\end{equation}
We have used the periodicity of $\chi_{(\ell,r)}$. 
Also, the fields $\psi_{(\ell,r)}e^{i(r+k)\phi/\sqrt{2k}}$  satisfy the same locality condition as $\psi_{(\ell,r)}e^{ir\phi/\sqrt{2k}}$, thus they are in the $\widehat{SU(2)}_{k}$ theory as well. In fact, they are the $k-\ell$ multiplet with character 
\begin{equation}
\frac{1}{\eta(\tau)}\sum_{t=0}^{k-1}\chi_{(\ell,\ell-2t)}\sum_{p=-\infty}^{\infty}q^{(\ell-2t+k+2pk)^{2}/4k}=\frac{1}{\eta(\tau)}\sum_{t=0}^{k-1}\chi_{(k-\ell,\ell-k+2t)}\sum_{p=-\infty}^{\infty}q^{(\ell-k+2t+2pk)^{2}/4k}.\end{equation}
 
The highest weight state in $\psi_{(\ell,r)}$ with $r=-\ell,\dots, \ell$ and $\ell=0,\dots, k$ has $h=\frac{\ell(\ell+2)}{4(k+2)}-\frac{r^{2}}{4k}$, which means for any integer $p$ the highest weight state in $\psi_{(\ell,\ell+2pk-2t)}$ has weight
\begin{equation}
h_{(\ell,l+2pk-2t)}=\left\{ \begin{array}{l@{\quad \mbox{if}\quad}c}
\frac{\ell(\ell+2)}{4(k+2)}-\frac{(\ell-2t)^{2}}{4k}&t=0,\dots, l\\
\frac{(k-\ell)(k-\ell+2)}{4(k+2)}-\frac{(\ell-2t+k)^{2}}{4k}
&t=\ell+1,\dots, k-1
\end{array}\right.\end{equation}

From the modular transformation property of $\widehat{SU(2)}_{k}$ and $\widehat{U(1)}$ we can also get that of the parafermions as follows
\begin{equation}
{\mathcal{S}_{(\ell,r)}}^{(\ell',r')}=\frac{2}{\sqrt{k(k+2)}}\sin\frac{\pi(\ell+1)(\ell'+1)}{k+2}e^{-i\pi  rr'/k}.
\end{equation}
Now, instead of $J_{\pm}$, let's use $J_{1+M/2}^+\sim\psi e^{i\sqrt{2/k+M}\phi}$, and $J_{1+M/2}^-\sim\psi^{\dagger} e^{-i\sqrt{2/k+M}\phi}$. The subscript indicates the weight of the currents. The same locality condition tells us that $\psi_{(\ell,r)}$ has to be multiplied by $\exp\left[i\frac{1}{\sqrt{k(kM+2)}}(kn+r)\phi\right]$ for any integer $n$. Together with the $J_{0}\sim i\partial \phi$ and other fields with $\ell=0$, the new currents form an extended chiral algebra. The character under this symmetry is now
\begin{equation}
\frac{1}{\eta(\tau)}\sum_{t=0}^{k-1}\chi_{(\ell,\ell-2t)}\sum_{p=-\infty}^{\infty}q^{\frac{[\ell-(Mk+2)t+kn+pk(Mk+2)]^{2}}{2k(Mk+2)}}.
\end{equation} 
The independent multiplets correspond to $n=0,\ldots, Mk+1$. When $M=0$, we get the original $\widehat{SU(2)}_{k}$ theory, as above. When $M\neq 0$, a little calculation shows that if $0\leq n\leq M\ell$, the highest weight state in the multiplet is $\psi_{(\ell,\ell-2[\frac{n}{M}])}e^{i\frac{\ell+nk-(Mk+2)[\frac{n}{M}]}{\sqrt{k(Mk+2)}}\phi}$, while if $M\ell+1\leq n\leq Mk+1$, the highest weight state in the multiplet is $\psi_{(\ell,\ell-2[\frac{n-1}{M}])}e^{i\frac{\ell+nk-(Mk+2)[\frac{n-1}{M}]}{\sqrt{k(Mk+2)}}\phi}$, where $[x]$ denotes the closest integer to $x$. We will call this character $\chi_{[\ell,r(\ell,n);s(\ell,n)]}$, where 
\begin{equation}
r(\ell,n)=\left\{ \begin{array}{l@{\quad \mbox{if}\quad}c}
\ell-2[\frac{n}{M}]&0\leq n\leq M\ell\\
\ell-2[\frac{n-1}{M}]&M\ell+1\leq n\leq Mk+1
\end{array}\right.\end{equation}
and
\begin{equation}
s(\ell,n)=\left\{ \begin{array}{l@{\quad \mbox{if}\quad}c}
\ell+nk-(Mk+2)[\frac{n}{M}]&0\leq n\leq M\ell\\
\ell+nk-(Mk+2)[\frac{n-1}{M}]&M\ell+1\leq n\leq Mk+1
\end{array}\right.\end{equation}
are the corresponding $U(1)$ charges of the primary fields.

Under modular S transformation, the standard Poisson resummation result tells us that
\begin{equation}\frac{1}{\eta(\tau)}\sum_{p=-\infty}^{\infty}q^{\frac{(r+pN)^{2}}{2N}}\to \sum_{s=0}^{N-1}\frac{1}{\sqrt{N}}e^{\frac{2\pi i rs}{N}}\frac{1}{\eta(\tau)}\sum_{p=-\infty}^{\infty}q^{\frac{(s+pN)^{2}}{2N}},\end{equation}
which gives the S transformation property of the combined theory as follows,
\begin{eqnarray*}
&&\chi_{[\ell,r(\ell,n);s(\ell,n)]}(-1/\tau)\nonumber\\
&=&\sum_{t=0}^{k-1}\sum_{\ell'=0}^{\frac{k+1}{2}}\sum_{t'=0}^{k-1}\frac{2}{k\sqrt{(k+2)(Mk+2)}}\sin\frac{\pi(\ell+1)(\ell'+1)}{k+2}\nonumber\\
&&\sum_{s=0}^{k(Mk+2)-1}e^{\frac{2\pi i (\ell-2t-Mkt+kn)s}{k(Mk+2)}-\frac{\pi i (\ell-2t)(\ell'-2t')}{k}}\frac{\chi_{(\ell',\ell'-2t')}(\tau)}{\eta(\tau)}\sum_{p=-\infty}^{\infty}q^{\frac{[s+pk(Mk+2)]^{2}}{2k(Mk+2)}}\nonumber\\
&=&\frac{2}{\sqrt{(k+2)(Mk+2)}}\sum_{\ell'=0}^{\frac{k+1}{2}}\sum_{n'=0}^{Mk+1}\sin\frac{\pi(\ell+1)(\ell'+1)}{k+2}\nonumber\\
&&\sum_{t'=0}^{k-1}e^{\frac{2\pi i (\ell+kn)(\ell'-2t'+n'k)}{k(Mk+2)}-\frac{\pi i \ell(\ell'-2t')}{k}}\frac{\chi_{(\ell',\ell'-2t')}}{\eta(\tau)}\sum_{p=-\infty}^{\infty}q^{\frac{[\ell'-2t'+n'k+pk(Mk+2)]^{2}}{2k(Mk+2)}}\nonumber\\
&=&\frac{2}{\sqrt{(k+2)(Mk+2)}}\sum_{\ell'=0}^{\frac{k+1}{2}}\sum_{n'=0}^{Mk+1}\sin\frac{\pi(\ell+1)(\ell'+1)}{k+2}e^{\frac{\pi i(-M\ell\ell'+2\ell n'+2\ell' n+2knn')}{Mk+2}}\chi_{[\ell',r(\ell',n');s(\ell',n')]}(\tau)
\end{eqnarray*}
i.e.,
\begin{equation}
{\mathcal{S}_{(\ell;n)}}^{(\ell';n')}=\frac{2}{\sqrt{(k+2)(Mk+2)}}\sin\frac{\pi(\ell+1)(\ell'+1)}{k+2}e^{\frac{\pi i(-M\ell\ell'+2\ell n'+2\ell' n+2knn')}{Mk+2}}.\label{eq:parasmatrix}\end{equation}
We have repeated used the periodicity of $\chi(\ell,r)$. To make the formula simpler, let's define $\sigma_{\ell}=\frac{1-(-1)^{\ell}}{2}$, thus since $k$ is odd, $\ell+\sigma_{\ell}k$ is always even. Define $n\to n-\sigma_{\ell}-\frac{(\ell+\sigma_{\ell}k)M}{2}$, then we can show that
\begin{equation}{\mathcal{S}_{(\ell;n)}}^{(\ell';n')}=\frac{2}{\sqrt{(k+2)(Mk+2)}}\sin\frac{\pi(\ell+1)(\ell'+1)}{k+2}(-1)^{\ell\ell'}e^{\frac{2\pi i k nn'}{Mk+2}}.\end{equation}
From the periodicity, we can still have $n=0,\dots, Mk+1$. Notice if we use the new parameter, the function of $U(1)$ charges of highest weight states in terms of $n$ has to be modified. The ${\cal S}$-matrix is now factorized into $\ell$ and $n$ parts. For $k=3$ and $M=1$ for example, this result coincides with that given in the text, eq. (\ref{k3M1}).

When $k$ is even, the loop with $\ell=k/2$ has only $k/2$ elements. The character of the multiplet in which the state $\psi_{(k/2,k/2)}e^{i\frac{1}{\sqrt{k(Mk+2)}}(kn+k/2)\phi}$ lives is
\begin{equation}\frac{1}{\eta(\tau)}\sum_{t=0}^{\frac{k}{2}-1}\chi_{(\frac{k}{2},\frac{k}{2}-2t)}\sum_{p=-\infty}^{\infty}q^{\frac{[\frac{k}{2}-(Mk+2)t+kn+p\frac{k(Mk+2)}{2}]^{2}}{2k(Mk+2)}}=\frac{1}{\eta(\tau)}\sum_{t=0}^{k-1}\chi_{(\frac{k}{2},\frac{k}{2}-2t)}\sum_{p=-\infty}^{\infty}q^{\frac{[\frac{k}{2}-(Mk+2)t+kn+pk(Mk+2)]^{2}}{2k(Mk+2)}}.\end{equation}
for $n=0,\dots, Mk/2$. Apart from the range diffence, the character looks just the same as before. So the calculation of the modular ${\cal S}$-matrix is almost the same as the odd $k$ case. 
\begin{eqnarray}
&&\chi_{[\ell;r(\ell,n);s(\ell,n)]}(-1/\tau)\nonumber\\
&=&\sum_{l'=0}^{\frac{k}{2}-1}\sum_{n'=0}^{Mk+1}{\mathcal{S}_{(\ell;n)}}^{(\ell';n')}\chi_{[\ell',r(\ell',n');s(\ell',n')]}(\tau)+\frac{2}{\sqrt{(k+2)(Mk+2)}}(\sum_{n'=0}^{\frac{Mk}{2}}+\sum_{n'=\frac{Mk}{2}+1}^{Mk+1})\sin\frac{\pi(\ell+1)(\frac{k}{2}+1)}{k+2}\nonumber\\
&&e^{\frac{\pi i(-M\ell\frac{k}{2}+2\ell n'+kn+2knn')}{Mk+2}}\sum_{t'=0}^{\frac{k}{2}-1}\frac{\chi_{(\frac{k}{2},\frac{k}{2}-2t')}(\tau)}{\eta(\tau)}\sum_{p=-\infty}^{\infty}q^{\frac{[\frac{k}{2}-2t'-Mkt'+n'k+pk(Mk+2)]^{2}}{2k(Mk+2)}}\nonumber\\
&=&\sum_{l'=0}^{\frac{k}{2}-1}\sum_{n'=0}^{Mk+1}{\mathcal{S}_{(\ell;n)}}^{(\ell';n')}\chi_{[\ell',r(\ell',n');s(\ell',n')]}(\tau)+\frac{2}{\sqrt{(k+2)(Mk+2)}}\sum_{n'=0}^{\frac{Mk}{2}}\frac{1+(-1)^{\ell}}{2}\nonumber\\
&&e^{\frac{\pi i(-M\ell\frac{k}{2}+2\ell n'+kn+2knn')}{Mk+2}}\left[\sum_{t'=0}^{\frac{k}{2}-1}+(-1)^{\ell+kn}\sum_{t'=\frac{k}{2}}^{k-1}\right]\frac{\chi_{(\frac{k}{2},\frac{k}{2}-2t')}(\tau)}{\eta(\tau)}\sum_{p=-\infty}^{\infty}q^{\frac{[\frac{k}{2}-2t'-Mkt'+n'k+pk(Mk+2)]^{2}}{2k(Mk+2)}}\nonumber\\
&=&\sum_{\ell'=0}^{\frac{k}{2}-1}\sum_{n'=0}^{Mk+1}{\mathcal{S}_{(\ell;n)}}^{(\ell';n')}\chi_{[\ell',r(\ell',n');s(\ell',n')]}(\tau)+\sum_{n'=0}^{\frac{Mk}{2}}{\mathcal{S}_{(\ell;n)}}^{(\frac{k}{2};n')}\chi_{[\frac{k}{2},r(\frac{k}{2},n');s(\frac{k}{2},n')]}(\tau)
\end{eqnarray}
We saw that the ${\cal S}$-matrix is the same as $k$ odd case. The only difference is that when $k$ is even, the $\ell$ blocks are not of the same size. To simplify, the best we can do is to make the redefinition $n\to n-\ell M/2$, then
\begin{equation}{\mathcal{S}_{(\ell;n)}}^{(\ell';n')}=\frac{2}{\sqrt{(k+2)(Mk+2)}}\sin\frac{\pi(\ell+1)(\ell'+1)}{k+2}i^{M\ell\ell'}(-1)^{n\ell'+\ell n'}e^{\frac{2\pi i k nn'}{Mk+2}}.\end{equation}

To summarize, the $\ZZ_{k}$-parafermions 
coupled with one $U(1)$, with symmetry generators $J_0\sim i\partial\phi$, $J_{1+M/2}^+\sim\psi e^{i\sqrt{M+\frac{2}{k}}\phi}$, and $J_{1+M/2}^-\sim\psi^{\dagger} e^{-i\sqrt{M+\frac{2}{k}}\phi}$, will have $\frac{(k+1)(Mk+2)}{2}$ multiplets. We can use the $(\ell;n)$ to label them, with $\ell$ the parafermion loop the state sits and together with $n$ determines the $U(1)$ charge assignment. When $k$ is odd (even), $\ell=0,\dots, \frac{k+1}{2}$ ($\ell=0,\dots, \frac{k}{2}+1$), $n=0,\dots, Mk+1$($n=0,\dots, \frac{Mk}{2} \mbox{\,if\,}\ell=\frac{k}{2}$). 


\providecommand{\href}[2]{#2}\begingroup\raggedright\endgroup

\end{document}